\documentclass{pazhastl}

\usepackage{apjfonts}

\usepackage{graphicx}
\usepackage{amsmath}
\usepackage{epsfig,graphics}
\usepackage{rotating} 
   
\newcommand {\be}{\begin{equation}}
\newcommand {\ee}{\end{equation}}

\begin{document}

\journalinfo{2011}{37}{5}{311}[331] 
 
\title{Vertical Structure of the Outer Accretion Disk in Persistent Low-Mass
  X-Ray Binaries} \author{\bf 
  A.V.~Mescheryakov\email{mesch@iki.rssi.ru}\address{1},
  N.I.~Shakura\address{2}, V.F.~Suleimanov\address{3,4} \addresstext{1}{\it
    Space Research Institute, Russian Academy of Sciences, Profsoyuznaya ul.
    84/32, Moscow, 117997 Russia} \addresstext{2}{\it Moscow State
    University, Moscow, 119991 Russia} \addresstext{3}{\it Kazan Federal
    University, Kremlevskaja ul. 18, Kazan, 420008 Russia}
  \addresstext{4}{\it Institute of Astronomy and Astrophysics, T\"{u} bingen
    University, Germany} } \shortauthor{Mescheryakov et al.}
  
\shorttitle{Vertical Structure of the Outer Accretion Disk in LMXB}

\submitted{September 6, 2010}

\begin{abstract}
  We have investigated the influence of X-ray irradiation on the vertical
  structure of the outer accretion disk in low-mass X-ray binaries by
  performing a self-consistent calculation of the vertical structure and
  X-ray radiation transfer in the disk. Penetrating deep into the disk, the
  field of scattered X-ray photons with energy $E\gtrsim10$\,keV exerts a
  significant influence on the vertical structure of the accretion disk at a
  distance $R\gtrsim10^{10}$\,cm from the neutron star. At a distance
  $R\sim10^{11}$\,cm, where the total surface density in the disk reaches
  $\Sigma_0\sim20$\,g\,cm$^{-2}$, X-ray heating affects all layers of an
  optically thick disk. The X-ray heating effect is enhanced significantly
  in the presence of an extended atmospheric layer with a temperature
  $T_{atm}\sim(2\div3)\times10^6$\,K above the accretion disk. We have
  derived simple analytic formulas for the disk heating by scattered X-ray
  photons using an approximate solution of the transfer equation by the
  Sobolev method. This approximation has a $\gtrsim10$\,\% accuracy in the
  range of X-ray photon energies $E<20$\,keV.

\medskip
DOI:  10.1134/S1063773711050045


\end{abstract}
 
\setcounter{page}{1}

 
\section{INTRODUCTION}
\label{sec:intro}
Low-mass X-ray binaries (LMXBs) consist of a neutron star or a black hole in
a pair with an optical star of mass $M\lesssim1M_\odot$. The companion star
in a close binary fills its Roche lobe and outflows onto a compact object
with the formation of an accretion disk.

The possibility of irradiation of the outer accretion disk in X-ray binaries
was first pointed out by Shakura~and~Sunyaev~(1973). Since the outer
accretion disk has a curved surface ($H\propto R^{n}$, $n=9/8$ for a
standard disk), the fraction of the X-ray flux from the central source that
is absorbed and ''thermalized'' near the disk surface depends on radius as
$Q_\star\propto1/R^2$. At the same time, the intrinsic energy release
(through viscosity) in the disk decreases with radius faster:
$Q_{vis}\propto1/R^3$. For a binary with a neutron star, the disk surface
heating through irradiation exceeds the intrinsic energy release already at
radii $R\gtrsim10^9$~cm.

The vertical structure of the outer accretion disk in LMXBs with direct
irradiation from the central source was investigated by many authors (see,
e.g., Tuchman et al. 1990; Vrtilek et al. 1990; Dubus et al. 1999). The
X-ray flux is commonly assumed to be thermalized near the accretion disk
photosphere (see, e.g., Dubus et al. 1999). In this case, external
irradiation has no significant influence on the vertical structure of the
disk as a whole (Lyutyi and Sunyaev 1976). Indeed, surface heating affects
the temperature in the central plane of the accretion disk if the following
condition is met:
\be 
\frac{Q_\star}{Q_{vis}}\gtrsim\tau_0 ~,
\label{uslovie_LS}
\ee 
where $\tau_0$ is the total optical depth of the disk in the vertical
direction for intrinsic radiation. The standard disk (Shakura and Sunyaev
1973) has a fairly large optical depth in the vertical direction
($\tau_0\gtrsim500$). The surface heating of such a disk cannot have a
significant effect on the conditions in the central plane up to very large
radii.

The model of a vertically isothermal accretion disk (Vrtilek et al. 1990) is
commonly used to describe LMXB observations. For example, the disk thickness
at the outer edge was estimated in this model from the observed amplitude of
optical variations on the light curves of LMXBs (de Jong et al. 1996),
$H/R\approx0.2$, which is a factor of $2\div3$ larger than the thickness of
an unirradiated standard disk (Shakura and Sunyaev 1973). Note that using
the simple model of an isothermal disk (Vrtilek et al. 1990) to describe
observations is inconsistent with the assumption about X-ray heating of only
the surface. As has been noted above, such heating has no significant effect
on the vertical disk structure, including the temperature profile.

The necessity of an increase in the accretion disk thickness at the outer
edge by a factor of $1.5\div2$ compared to the standard disk thickness also
follows from the modeling of the observed X-ray light curves (Esin et al.
2000; Suleimanov et al. 2008). Note that this excess thickness can be
explained by the presence of a hot atmosphere (see below) or a population of
relatively cold clouds embedded in the atmosphere (Suleimanov et al. 2003)
above the disk.

In an unirradiated accretion disk at a radius $R\approx5\times10^{10}$~cm,
the temperature on the photosphere drops below $10000$~K. As a result, a
zone with incomplete hydrogen ionization appears in the disk and thermal
instability must disrupt the stationary regime of accretion. On the other
hand, as follows from observations, the semimajor axis of the orbit in
persistent long-period LMXBs with known parameters (see, e.g., Table 1 in
Gilfanov and Arefyev 2005) can be $a\approx(2\div5)\times10^{12}$~cm (the
binaries GX 13+1 and Cyg X-2). The outer radius of the stationary accretion
disk in such binaries, $R_{out}\approx(2\div8)\times10^{11}$~cm , is
considerably larger than the radius at which a zone with incomplete hydrogen
ionization appears in an unirradiated disk.

The disk heating through X-ray irradiation, in principle, can increase the
disk temperature, thereby moving the region with incomplete hydrogen
ionization farther along the disk radius. However, as was shown previously
(see, e.g., Dubus et al. 1999), the central X-ray source cannot directly
irradiate the outer disk region where hydrogen transformed into a neutral
state, because the disk thickness in this zone decreases sharply (due to an
increase in themolecular weight of the material during hydrogen
recombination).  The problem with the screening of the outer accretion disk
can be resolved by assuming the presence of a scattering atmosphere with a
temperature $T_{atm}\approx(2\div3)10^6$~K above the disk.

X-ray irradiation changes greatly the structure of the near-surface layers
in an accretion disk. A thick hot ($T\sim(10^6\div10^7)$~K) plasma layer,
which we call an atmosphere, can be formed above its surface. The
calculations of a hot extended atmosphere applicable to the accretion disks
in LMXBs were performed in a number of works (Raymond 1993; Ko and Kallman
1994; Suleimanov et al. 1999; Jimenez-Garate et al. 2002).

As was shown by Jimenez-Garate et al. (2002), there exists a feedback
mechanism between the disk irradiation and the thickness of its atmospheric
layer, which leads to a significant increase of the latter.  The atmospheric
layer is optically thin in the vertical direction but optically thick in
radial coordinate.  Thus, for X-ray photons from the central source, the
thickness of the outer accretion disk increases by the height of the
atmosphere. For example, in the above paper, the half-thickness of a disk
with an atmosphere at $R=10^{11}$~cm was found to be
$z_{atm}/R\approx0.11\div0.20$ for an accretion rate in the range
$\dot{M}=(0.1\div1)\dot{M}_{\rm Edd}$, while the disk half-thickness at the
photospheric level was $z_{ph}/R\approx0.062\div0.083$ in the same range of
$\dot{M}$.  The presence of an extended atmosphere above the disk (as in
Jimenez-Garate et al. 2002) naturally explains the enhanced disk thickness
at the outer edge $H/R\approx0.2$ found by de Jong et al. (1996).

Here, we constructed a self-similar model of an irradiated accretion disk by
taking into account the scattering of X-ray photons in the disk. We also
took into account the possibility that an extended atmosphere was present
above the disk (we took the atmospheric parameters from Jimenez-Garate et
al. (2002), see \S2.1). We expect that the Xray photons after their
scattering in the disk and the atmosphere can penetrate and be thermalized
deeper in the disk than the direct photons from the central source. As will
be shown below, the scattering effect leads to deep heating of all disk
layers in z coordinate at a radius $R\sim10^{11}$~cm. In this case, the disk
still remains optically thick for intrinsic radiation.

The paper is organized as follows. In Section 2, we consider our model for
calculating the vertical structure of an irradiated disk that includes the
calculation of the transfer of X-ray photons in the disk and the atmosphere
by the Sobolev method (Sect. 2.2) and the determination of the vertical disk
structure under the photosphere (Sect. 2.3). The results of our calculation
of the vertical structure of an irradiated accretion disk and their
discussion are presented in Section 3. In the final section, we summarize
our conclusions.

\section{CALCULATING THE VERTICAL STRUCTURE OF THE OUTER ACCRETION DISK WITH
  IRRADIATION FROM A CENTRAL X-RAY SOURCE}
\label{sec:irr_disk_model}
 
\subsection{Irradiation Parameters}
\label{sec:irr_pars}
Consider a geometrically thin, optically thick accretion disk around a
neutron star of mass $M=1.4\,M_\odot$ in a LMXB. We assume the accretion
rate in the disk to be constant: $\dot{M} = const$. The chemical composition
of the disk corresponds to the cosmic elemental abundances: X=0.73, Y=0.25,
Z=0.017 (Allen 1973).

The X-ray radiation comes from a region near the compact object (a point
source for the outer disk) and has an energy spectrum $S(E)$ corresponding
to bremsstrahlung with a temperature $T_{sp}=8$~keV. In the energy range
$1\div20$~keV, the spectrum is
\be 
S(E) \propto \bigg(\frac{E}{k T_{sp}}\bigg)^{-0.4} \exp\bigg(-\cfrac{E}{k T}\bigg) 
\label{eq:Sp}
\ee
This model spectrum corresponds to the observed X-ray spectra of persistent
LMXBs that can be described by a bremsstrahlung spectrum with a temperature
$T_{sp}=5\div10$~keV (Liu et al. 2001). The disk--neutron star boundary
layer is probably the main source of the radiation illuminating the outer
accretion disk regions.

The X-ray luminosity of the central source is
\be 
L_X= \epsilon_X \dot{M}\,c^2 , 
\ee
where $\dot{M}$ is the mass accretion rate, the accretion efficiency onto
the neutron star is assumed to be $\epsilon_X\approx0.1$.

An X-ray flux (the central source is assumed to be a pointlike and
isotropically emitting one),
\be
F^\nu_X(\nu)=\frac{L_X}{4 \pi R^2} S(\nu) .
\label{eq:Fx}
\ee 
is incident on the outer disk at radius R. Note that Eq. (4) is approximate.
Thus, assuming the Xray radiation to come from a boundary layer with a
thickness much smaller than the neutron star radius without relativistic
effects, $2\pi^2$ must appear in the denominator of Eq. (4) instead of the
coefficient $4\pi$.

\begin{figure}
  \centering
  \includegraphics[width=1.0\linewidth]{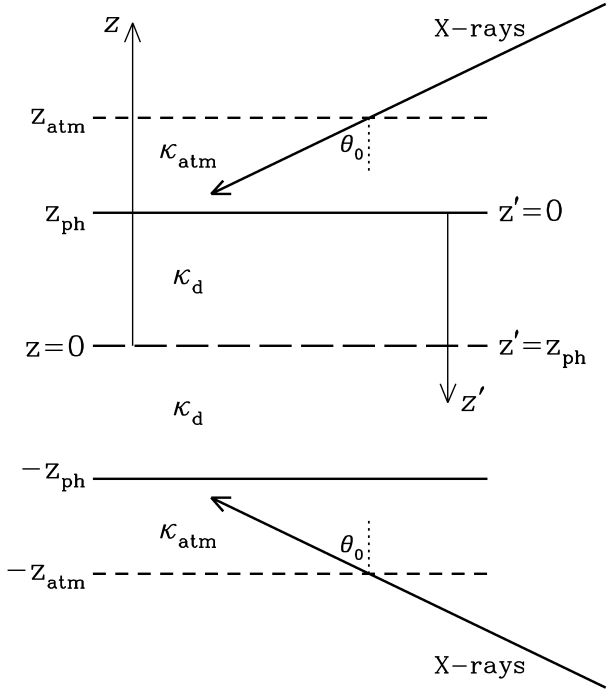}
  \caption{Irradiation geometry of a disk with an atmosphere and the
    coordinate system in the disk.}
  \label{fig:slab3zone} 
\end{figure}

The angle between the direction of incidence of the X-ray photons and the
inward normal to the disk/atmospheric layer surface (see Fig. 1) is
\be
\theta_0 = \frac{\pi}{2} -
arctg\bigg(\frac{d\,H}{d\,R}\bigg) + arctg\bigg(\frac{H}{R}\bigg) ,
\ee
where the function $H(R)$ represents the dependence of the disk
half-thickness (or the atmosphere height) on radius for X-ray photons.

If the shape of the geometrically thin ($H/R\ll\,1$) disk/atmosphere can be
described by a power-law dependence on radius $H(R){\propto}R^n$, then the
irradiation parameter $\zeta_0=\cos\Theta_0$ is expressed as
\be 
\zeta_0 \approx \frac{d H}{d R}-\frac{H}{R} = (n-1)\frac{H}{R} ~.
\label{eq:zeta0}
\ee

We considered two models for the irradiation of the outer accretion disk:
with and without an atmospheric layer above the disk.

\paragraph{(1) The irradiation of a disk without an atmosphere.}
In this case, the disk thickness for X-ray photons corresponds to the
photospheric level $H=z_{ph}$ for intrinsic radiation. The irradiation
parameter $\zeta_0$ depends on disk thickness and index $n_{ph}$ (see Eq.
(6)).  We found zph simultaneously with the solution of the equations for
the vertical disk structure (see below).  We fixed the power-law slope of
the surface at $n_{ph}=9/8$ (which corresponds to the outer parts of an
unirradiated standard disk).

\paragraph{(2) The irradiation of a disk with an atmosphere.}
There is an atmospheric layer that is optically thin in the vertical
direction ($\tau\ll1$), but optically thick in the radial direction above
the disk photosphere (for the possibility of the appearance of such a layer
above the outer accretion disk, see Jimenez-Garate et al. 2002).

A detailed calculation of the structure of an extended layer above an
accretion disk is a nontrivial problem that is beyond the scope of this
paper. To specify the parameters of the atmosphere above the disk, we used
the results of atmospheric-layer calculations from Jimenez-Garate et al.
(2002) obtained with similar irradiation parameters and chemical composition
of the accretion disk. In the above paper, the atmosphere was assumed to be
in hydrostatic equilibrium. The stationary model atmosphere is applicable
for most LMXBs with X-ray emission lines from a photoionized plasma observed
in their spectra (X-ray pulsars and X-ray binaries with orbital inclinations
$i>70^\circ$), except for binaries with a strong outflow of matter from the
accretion disk (for more detail, see the discussion in \S8.1 of
Jimenez-Garate et al. (2002)).

The disk thickness for X-ray photons corresponds to the photospheric level
$H=z_{atm}$. We assumed $z_{atm}$ and the slope of the surface of the
atmospheric layer natm at a given radius to be (see Jimenez-Garate et al.
2001)
\be  
z_{atm}/R = \left[
  \begin{array}{cc}
    1.2\times10^{-3} R^{0.18} &, L/L_{\rm Edd}=0.1 \\
    1.0\times10^{-3} R^{0.21} &, L/L_{\rm Edd}=1.0 ~.
  \end{array}
  \right.
\ee
The total surface density in the atmospheric layer above the disk (see Fig.
14 in Jimenez-Garate et al. 2002) is
\be
\Sigma_{atm} [g/cm^2]\approx \left[
  \begin{array}{cc}
    0.066 &, L/L_{\rm Edd}=0.1 \\
    0.32 &, L/L_{\rm Edd}=1.0
  \end{array}
  \right.
\label{eq:Satm}
\ee

According to the calculations by Jimenez-Garate et al. (2002), the bulk of
the surface density in the atmospheric layer is accumulated in a zone with a
temperature $T\approx2\div3\times10^6$~K, where the balance between heating
and cooling is maintained by the photoionization/recombination of
hydrogenand helium-like ions. A more rarefied region with a temperature
$T\gtrsim2\times10^7$~K, where the Compton heating and cooling processes
dominate, lies higher.  Its contribution to $\Sigma_{atm}$ is, on average,
$\sim15\%$ at radii $R=10^{10}\div10^{11}$~cm. As follows from the cited
calculation, the total surface density in the atmospheric layer
$\Sigma_{atm}$ is virtually independent of the radius in this range of $R$.

In our simplified model of the atmospheric layer, we assumed its temperature
to be $T_{atm}\approx2\times10^6$\,K.  The pressure near the base of an
isothermal atmosphere is
\be \nonumber
P_{atm} = \Sigma_{atm} \bigg(\frac{2 \Re T_{atm}}{\pi\mu_{atm}}\bigg)^{1/2}
\Omega_K \Phi\bigg(\frac{z_s}{z_0}\bigg) ~,
\ee 
where $\Re$ is the universal gas constant, $\Omega_K=\sqrt{G M / R^3}$ is
the angular Keplerian velocity at radius $R$, $\mu_{atm}$ is the molecular
weight of the material, $z_s$ is the height near the base of the atmospheric
layer, $z_0=\sqrt{\frac{2\Re T_{atm}}{\Omega_K^2 \mu_{atm}}}$, and the
function $\Phi(x)$ is
\be
\nonumber
\Phi(x) = \frac{\pi}{2}\cfrac{e^{-x^2}}{\int_x^\infty e^{-t^2} dt} ~.
\ee
Since the height near the base of the atmospheric layer $z_s\approx z_{ph}$,
we obtain $\frac{z_s}{z_0}\lesssim1$ and $\Phi(z_s/z_0)\approx1$.  Thus, the
formula for the pressure near the base of the atmosphere takes the form
\be
P_{atm} \approx \Sigma_{atm} \bigg(\frac{2\Re T_{atm}}{\pi\mu_{atm}}\bigg)^{1/2} \Omega_K ~.
\label{eq:Patm}
\ee
The molecular weight of the material in the atmospheric layer (a completely
ionized medium) is $\mu_{atm}=0.6$.

\subsection{Calculating the X-ray Radiation Transfer in the Disk and the
  Atmospheric Layer in the Sobolev Approximation}
\label{sec:Xrays_RT}

Being absorbed and thermalized in subphotospheric layers, the X-ray
radiation can serve as an additional disk heating source.

To determine the mean intensity and flux of Xray photons in the accretion
disk, we solved the onedimensional problem of X-ray photon transfer in a
layer of finite thickness in the Sobolev approximation (see the Appendix).
The following simplifications were used:
\begin{enumerate}
\item The scattering in the medium was assumed to be coherent (Thomson
  scattering). This approximationmay be considered justified, because the
  radiation incident on the disk has a fairly soft spectrum. The effect
  fromthe change in the frequency of photons due to the direct Compton
  effect as they are scattered in the disk will be small for photons with
  energy $E<20$~keV (for more detail, see below).

\item In solving the transfer equation, we use the Eddington approximation
  (see Eq. (A14)).

\item We divided a plane-parallel layer of finite thickness into three zones
  in each of which the absorption coefficient depends only on frequency:
  $\kappa=\kappa(\nu)$. The central zone corresponds to a cold disk
  ($\kappa=\kappa_d$), while the zones at the top and the bottom describe
  the atmospheric layer on both sides of the disk ($\kappa=\kappa_{atm}$)
  (see Fig. 1).

  We assume that the opacity in the disk is determined by photoabsorption
  for a cold gas (Morrison and McCammon 1983):
  \be
  \kappa_d = \kappa_{MM}(\nu) .
  \ee
  For the atmosphere, we will consider two cases:
  \begin{itemize}
  \item a completely scattering atmosphere:
    \be
    \kappa_{atm}(\nu)=0 .
    \label{eq:katm1}
    \ee
  \item the absorption and scattering coefficients in the atmosphere are
    approximately equal:
    \be
      \kappa_{atm}= \left[
        \begin{array}{cc}
          \sigma &, \kappa_{MM}>\sigma \\
          \kappa_{MM} &, \kappa_{MM}\le\sigma ,
        \end{array}
      \right.
      \label{eq:katm2}
      \ee
      where $\sigma$ is the scattering coefficient.
  \end{itemize}
\end{enumerate} 

Solving the transfer equation for X-ray photons by the Sobolev method, we
obtained fairly simple analytic formulas for the mean intensity and flux of
X-ray photons at a given frequency as a function of the surface density
(which is accumulated from the surface deep into the disk) $\Sigma$ and
parameters $\Sigma_0$, $\Sigma_{atm}$, as well as the specified X-ray
opacities $\kappa^\nu_d$ and $\kappa^\nu_{atm}$:
\begin{eqnarray}
J^\nu_{tot} = J^\nu_{tot}(\Sigma;\Sigma_0,\Sigma_{atm},\kappa^\nu_d,\kappa^\nu_{atm}) , \\
H^\nu_{tot} = H^\nu_{tot}(\Sigma;\Sigma_0,\Sigma_{atm},\kappa^\nu_d,\kappa^\nu_{atm}) ,
\end{eqnarray}
where $\Sigma_0$ is the total surface density in the disk.  Here and below,
the superscript $\nu$ emphasizes the dependence of a quantity on frequency.

The mean intensity and flux of X-ray photons in the disk at frequency $\nu$
are (for more detail, see Appendix 1)
\begin{eqnarray}
  \nonumber 
  J^\nu_{tot}(\Sigma) = 
  \cfrac{F^\nu_X}{4\pi}\bigg\{ 
C^\nu\bigg[e^{-k\tau^\nu} + 
   e^{-k(\tau^\nu_0-\tau^\nu)}\bigg] + \\
 +(1-D^\nu)
\bigg[ e^{-\tau^\nu/\zeta_0} +
e^{-(\tau^\nu_0 - \tau^\nu)/\zeta_0} \bigg] \bigg\} , \label{eq:Jtot_disk}\\
  \nonumber 
  H^\nu_{tot}(\Sigma) = 
  F^\nu_X\bigg\{ 
\cfrac{k\,C^\nu}{3}\bigg[e^{-k\tau^\nu} - 
    e^{-k(\tau^\nu_0-\tau^\nu)}\bigg] + \\ 
 +\bigg(\zeta_0-\cfrac{D^\nu}{3\,\zeta_0}\bigg)
\bigg[ e^{-\tau^\nu/\zeta_0} -
e^{-(\tau^\nu_0 - \tau^\nu)/\zeta_0} \bigg] \bigg\} ~,
  \label{eq:Htot_disk}  
\end{eqnarray}
where $\tau^\nu_0$ is the total optical depth of the disk in the vertical
direction for X-ray radiation at frequency $\nu$,
$\tau^\nu=(\sigma+\kappa^\nu_d)\Sigma$,
$\lambda=\frac{\sigma}{\sigma+\kappa^\nu_d}$, $\kappa^\nu_d$ is the
absorption coefficient for X-ray photons with frequency $\nu$ in the disk,
$\sigma$ is the scattering coefficient, $k=\sqrt{3(1-\lambda)}$.

For a disk without an extended atmosphere, the coefficients $D^\nu$ and
$C^\nu$ are defined by Eqs. (A23) and (A27), respectively. For a disk with
an atmosphere, the coefficient $D^\nu$, as above, is calculated from Eq.
(A23), while the coefficient $C^\nu$ has a slightly more complex form (see
(A49)). Note that $D^\nu$ is not an independent coefficient; it is related
to $C^\nu$ (see, e.g., Eq. (A27)).

The additional heating of the disk by X-ray photons of a given frequency
$\epsilon^\nu$ is proportional to their mean intensity:
\be
\epsilon^\nu = 4\pi\rho\kappa^\nu_d J^\nu_{tot}.
\ee
Note that $J^\nu_{tot}$ tot includes both primary and scattered Xray
photons.

The local energy release in the disk through its irradiation by X-ray
photons with a spectrum $S(\nu)$ is
\be
\epsilon = \int_0^\infty \epsilon^\nu d\nu = 
4\pi\rho \int_0^\infty \kappa^\nu_d J^\nu_{tot} d\nu ~.
\label{eq:nagrev_irr}
\ee 
The total heating of the disk from the photosphere to the central plane
through its irradiation is
\be
Q_{irr}(\Sigma_{ph}) = \int_0^\infty H^\nu_{tot}(\Sigma_{ph}) d\nu ,
\label{eq:sum_nagrev_irr}
\ee
where $\Sigma_{ph}$ is the surface density in the layers above the disk
photosphere.

\paragraph{The accuracy of the approximate solution of the transfer equation
  by the Sobolev method when calculating the field of X-ray photons in the
  disk.}
It is necessary to estimate the accuracy of the simple formulas for the flux
and mean intensity of X-ray photons in the disk (17) and (18) derived by the
Sobolev method. For this purpose, we compared the plane albedos of a
semi-infinite layer $A_{Sob}$ (Eq. (A39)) obtained by the Sobolev method for
various incident X-ray photon energies (we used the energy dependence of the
absorption coefficient $\kappa_{MM}(E)$ for a cold gas) with the exact albedos:
\begin{itemize}
\item $A_{exact}$ --- the exact albedos for coherent scattering with the
  Rayleigh phase function were calculated via the Chandrasekhar $H$-functions
  (for more detail, see Appendix 3).
\item $A^{comp}_{sim}$ --- the albedos for scattering with the Compton
  effect. To calculate $A^{comp}_{sim}$, we used the method of Monte Carlo
  numerical simulations (see, e.g., Pozdnyakov et al. 1983). The
  semi-infinite layer was assumed to be cold, $k T\ll m_e c^2$; $10^7$ trial
  photons were taken for each albedo. To estimate the accuracy of our Monte
  Carlo simulations, we also determined the albedo by this method for
  coherent scattering $A_{sim}$, which can be compared with its exact value
  of $A_{exact}$.
\end{itemize}

 \begin{figure*}
   \centering
   \includegraphics[width=0.49\linewidth]{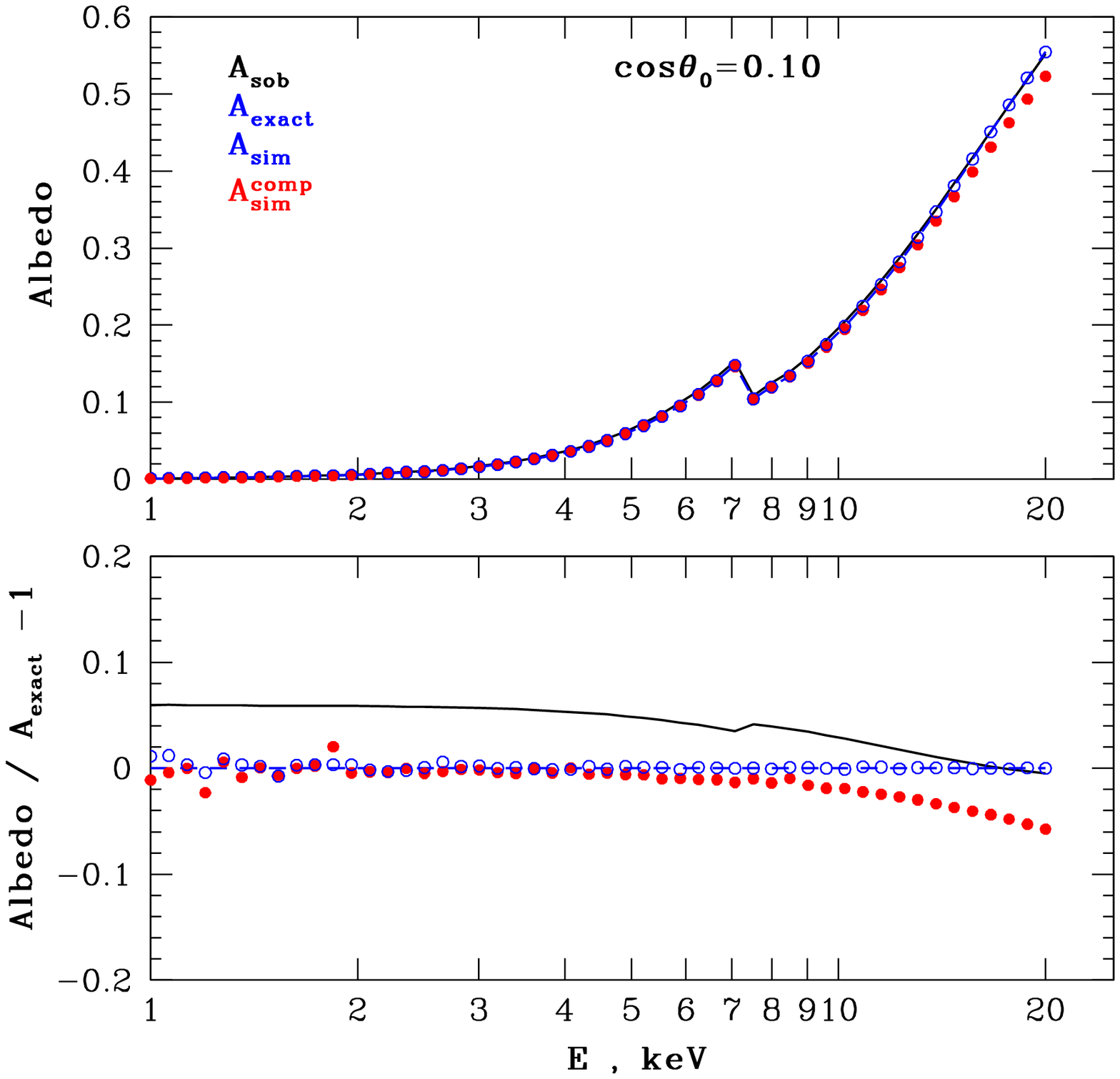}
   \includegraphics[width=0.49\linewidth]{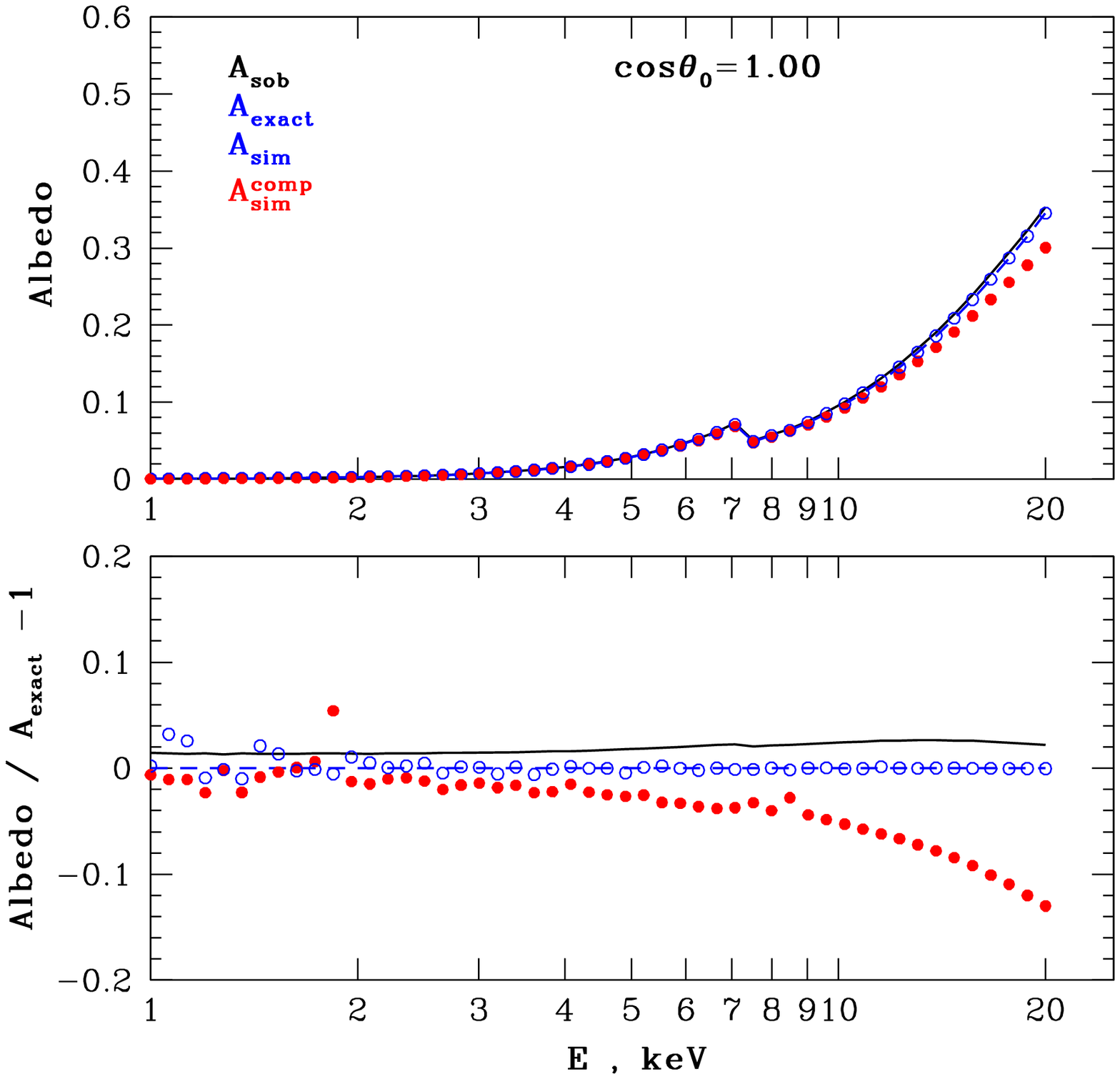}
   \caption{Albedo of a semi-infinite layer with the opacity of a cold gas
     $\kappa_{MM}$ versus energy $E$ for two angles of incidence of the
     X-ray photons on the layer: $\cos\theta=0.1$ (left panels) and
     $\cos\theta=1.0$ (right panels). The upper plots show: $A_{sob}$ (solid
     line)-the approximate Sobolev value, $A_{exact}$ (dashed line) --- the
     exact value for coherent scattering, $A_{sim}$ (open circles) --- from
     our numerical simulation for coherent scattering, $A^{comp}_{sim}$
     (filled circles) --- from our numerical simulation with the Compton
     effect. The lower plots show the ratios $A_{sob}/A_{exact}$,
     $A_{sim}/A_{exact}$ and $A^{comp}_{sim}/A_{exact}$.}
   \label{fig:compareA}     
 \end{figure*}

Figure 2 shows the dependence of various albedos ($A_{sob}$, $A_{exact}$,
$A_{sim}$, $A^{comp}_{sim}$) on energy $E$ (upper plots) and ratios
$A_{sob}/A_{exact}$, $A_{sim}/A_{exact}$, and $A^{comp}_{sim}/A_{exact}$
(lower plots) for two angles of incidence of the X-ray photons on the disk,
$\cos\theta_0=0.1$ (left) and $\cos\theta_0=1.0$ (right). We can estimate
the accuracy of the derived approximate formulas (17) and (18) for the mean
intensity and flux of X-ray photons in the disk as $\lesssim10\%$ for
photons with energy $E<20$~keV.

Note that calculating the field of X-ray photons with a higher accuracy will
require solving the problem of X-ray photon transfer in the disk by taking
into account the frequency change in each scattering.  This problem can be
solved numerically (for a cold, geometrically thin disk, see, e.g., Psaltis
2002).

\subsection{The System of Equations for the Vertical Structure of a Disk
  Irradiated by a Central X-ray Source and its Solution}
\label{sec:disk_vstruct}

The vertical structure of the accretion disk at a given radius can be
calculated under the assumption of a standard $\alpha$-disk (Shakura and
Sunyaev 1973); the $(r,\phi)$ component of the viscous stress tensor in the
disk is proportional to the pressure $w_{r\phi}={\alpha}P$.  Here, we
assumed that $\alpha=0.5$.

Denote the density, temperature, energy flux in the vertical direction
toward the surface, surface density, (Rosseland) optical depth and opacity,
and the photospheric level for intrinsic radiation (from the disk midplane)
by $\rho$, $T$, $Q$, $\Sigma$, $\tau^r$, $\kappa^r$, and $z_{ph}$,
respectively.  Let us introduce the vertical coordinate $z^{\prime}$
measured from the photospheric level deep into the disk (see Fig. 1). The
surface density $\Sigma$ is also measured toward the central plane of the
accretion disk:
\be 
\Sigma = \Sigma_{ph} + \int_0^{z^\prime}\rho d z, 
\ee 
where $\Sigma_{ph}$ is the surface density in a column of material above the
disk photosphere; the total surface density is $\Sigma_0$ in the disk and
$\Sigma=\Sigma_0/2$ in the its central plane.

To solve the one-dimensional problem of the vertical accretion disk
structure at a given radius ($R$), we sought a numerical solution of the
following system of ordinary differential equations (see also Suleimanov et
al. 2007):
\begin{enumerate}
\item The equation of hydrostatic equilibrium in $z$ coordinate:
\be
\label{eq:dPdz}
  \cfrac{d P}{d z^\prime} = \rho\Omega_K^2 (z_{ph} -z^\prime) ~.
\ee 
The contribution from the radiation pressure is insignificant in the outer
disk: $P_{gas}\gg\,P_{rad}$. We neglect the contribution from $P_{rad}$ to
the total pressure by assuming that $P=P_{gas}$.

\item The equation of energy transfer by radiation in $z$ coordinate (in the
  diffusion approximation):
\be
\label{eq:dTdz}
  \cfrac{dT}{d z^\prime} = \cfrac{3 \kappa^r \rho}{4 a c T^3} ~Q .
\ee
Note that the Rosseland approximation (24) is applicable only for fairly
deep layers, $\tau^r\gtrsim1$, under the disk photosphere. As long as the
accretion disk is optically thick in the vertical direction
($\tau_0^r\gg1$), using the diffusion approximation to determine the
vertical disk structure is justified. At the same time, to find, for
instance, the radiation spectrum from the accretion disk, the radiative
transfer in the near-surface layers where the emergent radiation is formed
must be calculated accurately (see, e.g., Suleimanov et al. 1999).

It should also be noted that Eq. (24) disregards the energy transfer through
convection. The convection mechanism is switched on when the vertical
temperature gradient (24) is larger than the adiabatic $\cfrac{dT}{d
  z^\prime} > \cfrac{dT}{dz^\prime}\bigg|_{ad}$.  The convection zone
emerges in the outer accretion disk where the temperature drops below
$T\sim10^4$\,K and hydrogen transforms into a neutral state (see, e.g.,
Meyer and Meyer-Hofmeister 1982). A thermal instability develops in the
zone with partial hydrogen ionization in the accretion disk. Since we
consider only the case of stationary accretion, we calculated the disk
structure up to the radius $R_K$ at which the zone with incomplete hydrogen
ionization appears.

\item The equation for energy release:
\be
\label{eq:dQdz}
  \cfrac{d Q}{d z^\prime} = -\cfrac{3}{2}\alpha  P \Omega_K  
- \epsilon ~.
\ee
The sum of two terms appears on the right-hand side of Eq. (25). The first
term represents the intrinsic energy release in a standard Shakura-Sunyaev
disk, while the second term represents the additional energy release due to
the absorption and thermalization of direct and scattered X-ray photons in
the accretion disk (without irradiation $\epsilon=0$). To find $\epsilon$
dependent on depth $z^\prime$, we use an approximate solution of the X-ray
radiation transfer problem (see Eq. (20)).

\item The equation for the surface density $\Sigma$:
\be
\label{eq:dSdz}
  \cfrac{d\Sigma}{d z^\prime} = \rho ~.
\ee

\item We find the optical depth $\tau^r$ from the equation:
\be
\label{eq:dtaudz}
\cfrac{d\tau^r}{d z^\prime} = \kappa^r \rho  ~. 
\ee
The Rosseland opacity in Eqs. (24) and (27) depends on density and
temperature: $\kappa^r=\kappa^r(\rho,T)$. For a given chemical composition,
we determined $\kappa^r$ using the Opacity Project tables (Badnell et al.
2005).

\end{enumerate}

The equation of state (for an ideal gas) should be
added to Eqs. (23)--(27):
\be
\label{eq:eqofstate}
 P = \cfrac{\Re}{\mu} \rho T .
\ee
We determined the molecular weight $\mu$ at the photospheric level of the
disk using the Saha formula (for a given chemical composition):
\be 
 \mu = \mu(P_{ph},T_{ph})  ~.
\label{eq:mu}
\ee
We found the radius $R_K$ at which the zone with incomplete hydrogen
ionization appeared in the accretion disk from the condition
\be
\mu(R_K) > \cfrac{\mu_{iH} + \mu_{ni}}{2} ,
\label{eq:Rk}
\ee
where $\mu_{iH}=0.66$ and $\mu_{ni}=1.26$ are the molecular weights of the
material in which only hydrogen is ionized (at the chemical composition
specified above) and a completely neutral medium, respectively. We
calculated the vertical structure of a stationary disk up to the radius
$R_K$.
 
To solve the system of ordinary differential equations (23)--(29) with (28)
and (29), we specified the following boundary conditions at the photospheric
level of the disk:
\be 
\left[
\begin{array}{lll}
P_{ph}      & = & \cfrac{\Omega_K^2
  z_{ph}}{\kappa^r(\rho_{ph},T_{ph})}\times\tau^r_{ph} + P_{atm}\\
T_{ph}      & = & \bigg( \cfrac{Q_{ph}}{\sigma_{SB}} \bigg)^{1/4} \\
Q_{ph}      & = & Q_{vis}  + Q_{irr}(\Sigma_{ph})\\
\Sigma_{ph} & = & \cfrac{\tau^r_{ph}}{\kappa^r(\rho_{ph},T_{ph})} + \Sigma_{atm}\\
\tau^r_{ph} & = & \cfrac{2}{3} 
\end{array}
\right. 
\label{eq:vstruct_bc}
\ee 
where $\rho_{ph}=\cfrac{\mu P_{ph}}{\Re T_{ph}}$; $\Sigma_{atm}$, $P_{atm}$,
$Q_{vis}$, $Q_{irr}$ are, respectively, the surface density and pressure of
the atmospheric layer (see Eqs. (8) and (10)), the total energy release in
the disk through viscosity, and the total energy release through X-ray
heating of the disk; $\sigma_{SB}$ is the Stefan-Boltzmann constant. We
derived the dependence $Q_{irr}(\Sigma)$ from the solution of the X-ray
radiation transfer problem (see (21)). We have $Q_{irr}=0$,
$\Sigma_{atm}=0$, for an unirradiated disk and $Q_{irr}>0$, $\Sigma_{atm}=0$
for an irradiated disk without an atmosphere.

The intrinsic energy release in the accretion disk at radius $R$ is defined by
the expression
\be
Q_{vis} = \frac{3\,G\,M\dot{M}}{8\pi\,R^3} f(R) ~,
\ee
where the function $f(R)$ is related to the boundary condition at the inner
accretion disk boundary. We investigate the outer disk where it can be
assumed that $f(R)=1$.

Given the boundary conditions (31), we integrated the system of equations
(23)--(27) from the photospheric level ($z^\prime=0$) deep into the disk
surface using the Runge-Kutta method implemented in the Numerical Recipes
software package (Press et al. 1992).  The height of the photosphere
$z_{ph}$ is a free parameter of the problem and is not known in advance. For
an irradiated disk, the total surface density in the disk $\Sigma_0$ is
another free parameter. We found $z_{ph}$ and $\Sigma_0$ when constructing
the solution by the method of successive iterations using an additional
condition for the energy flux in the central plane of the disk:
\be
\label{eq:Q_z=0}
Q\bigg|_{z^\prime=z_{ph}} = 0 .
\ee

\section{RESULTS AND DISCUSSION} 
\label{sec:results}

 \begin{figure*} 
   \centering 
   \includegraphics[width=1.0\linewidth]{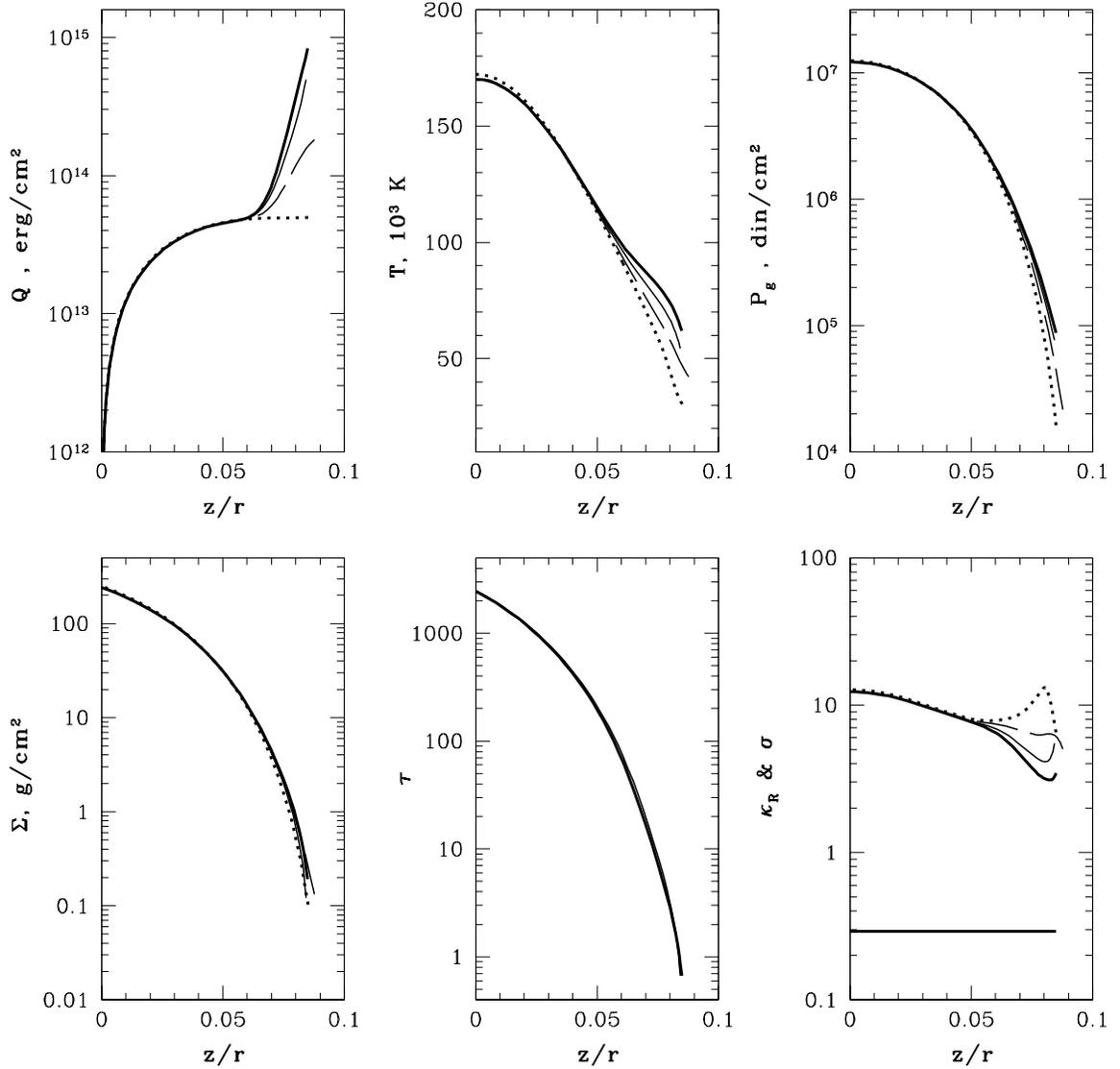}
   \caption{Vertical structure of an irradiated accretion disk:
     $R=10^{10}$~cm, $L_X=L_{\rm Edd}$. The central object is a neutron star
     with mass $M_1=1.4\,M_\odot$, the viscosity parameter in the disk is
     $\alpha=0.5$.  The thick solid line indicates the irradiation of a disk
     with a completely scattering atmosphere, the thin solid line indicates
     the irradiation of a disk with an atmosphere where absorption is equal
     to scattering for photons with energy $E\lesssim10$~keV, the dashed
     line indicates the irradiation of a disk without an atmosphere, and the
     dotted line indicates an unirradiated disk.}
   \label{fig:risAvstrIrr}  
\end{figure*}

 \begin{figure*}
   \centering 
    \includegraphics[width=1.0\linewidth]{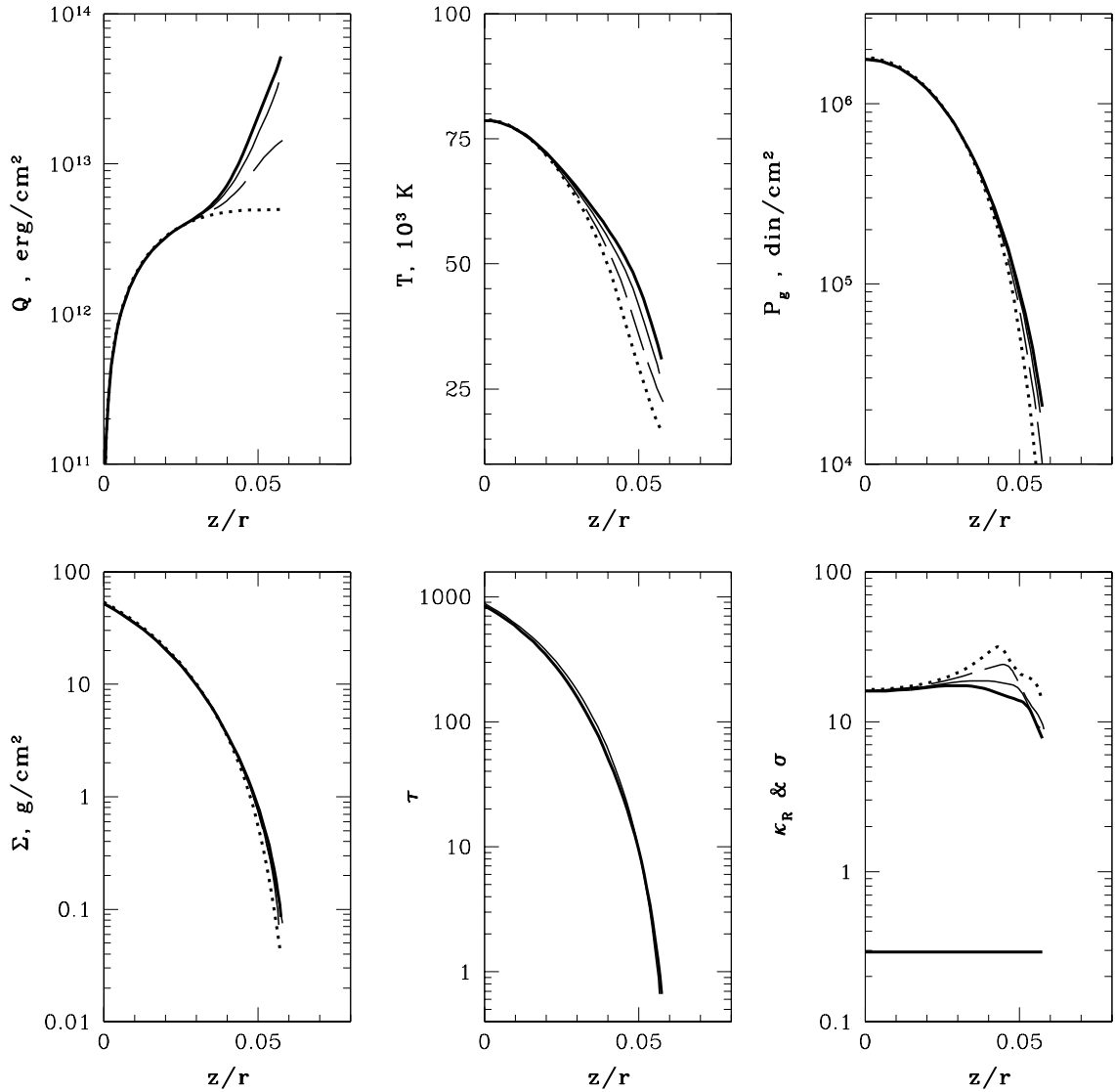}
    \caption{Vertical structure of an irradiated disk: $R=10^{10}$~cm,
      $L_X=0.1\,L_{\rm Edd}$. The designations of the lines for
      variousmodels of an irradiated disk are the same as those in Fig. 3.}
    \label{fig:risBvstrIrr}  
 \end{figure*}

 \begin{figure*}
   \centering 
    \includegraphics[width=1.0\linewidth]{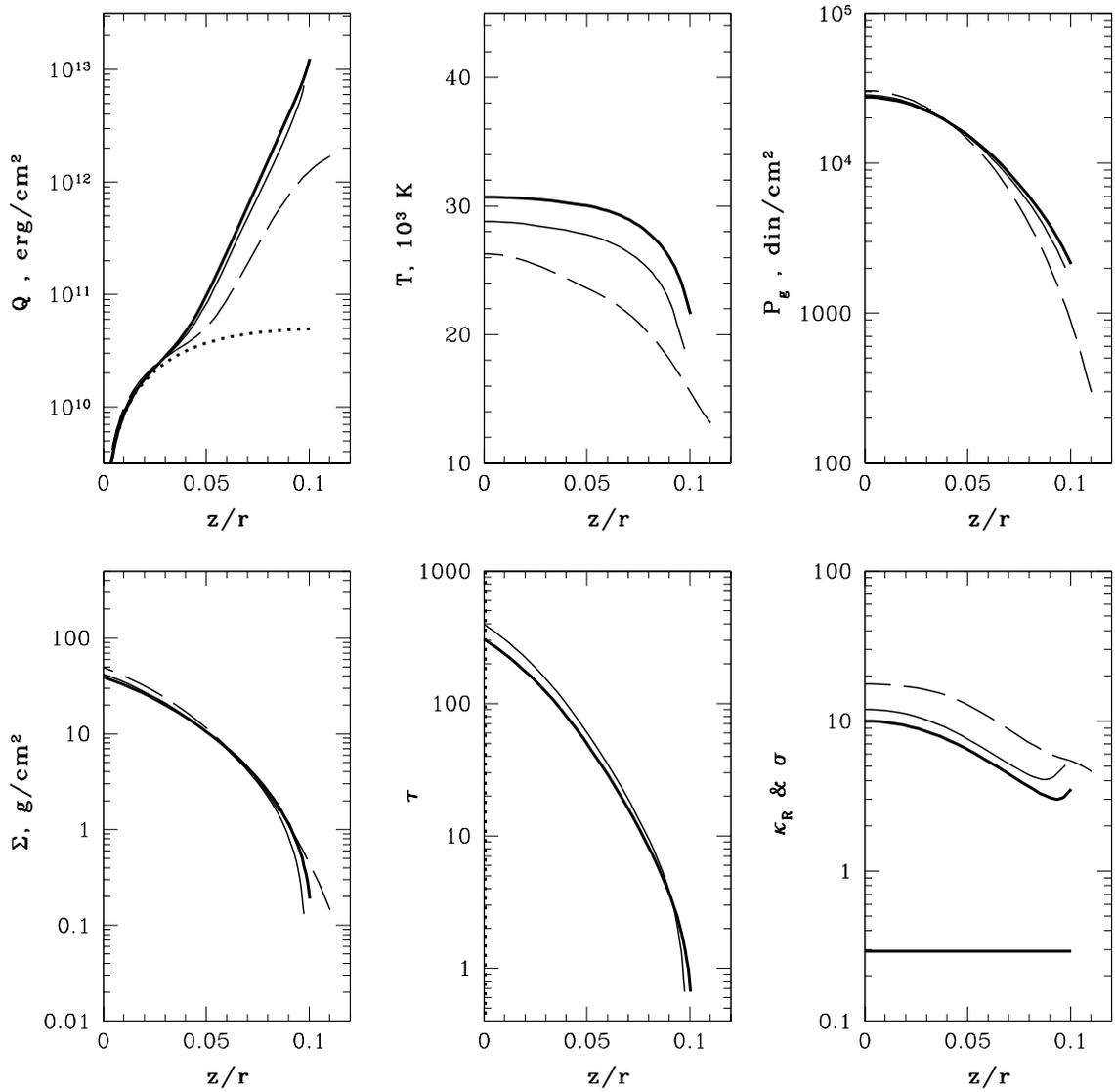}
    \caption{Vertical structure of an irradiated disk: $R=10^{11}$~см,
      $L_X=L_{\rm Edd}$. The designations of the lines for variousmodels of
      an irradiated disk are the same as those in Fig. 3. The model of an
      unirradiated disk is not shown. Instead, the dotted line on the plot
      of $Q(z)$ indicates the profile of the energy flux due to viscous
      heating ($Q_{vis}$) in the model of an irradiated disk (a completely
      scattering atmosphere).}
   \label{fig:risCvstrIrr}  
 \end{figure*}

 \begin{figure*}
   \centering 
    \includegraphics[width=1.0\linewidth]{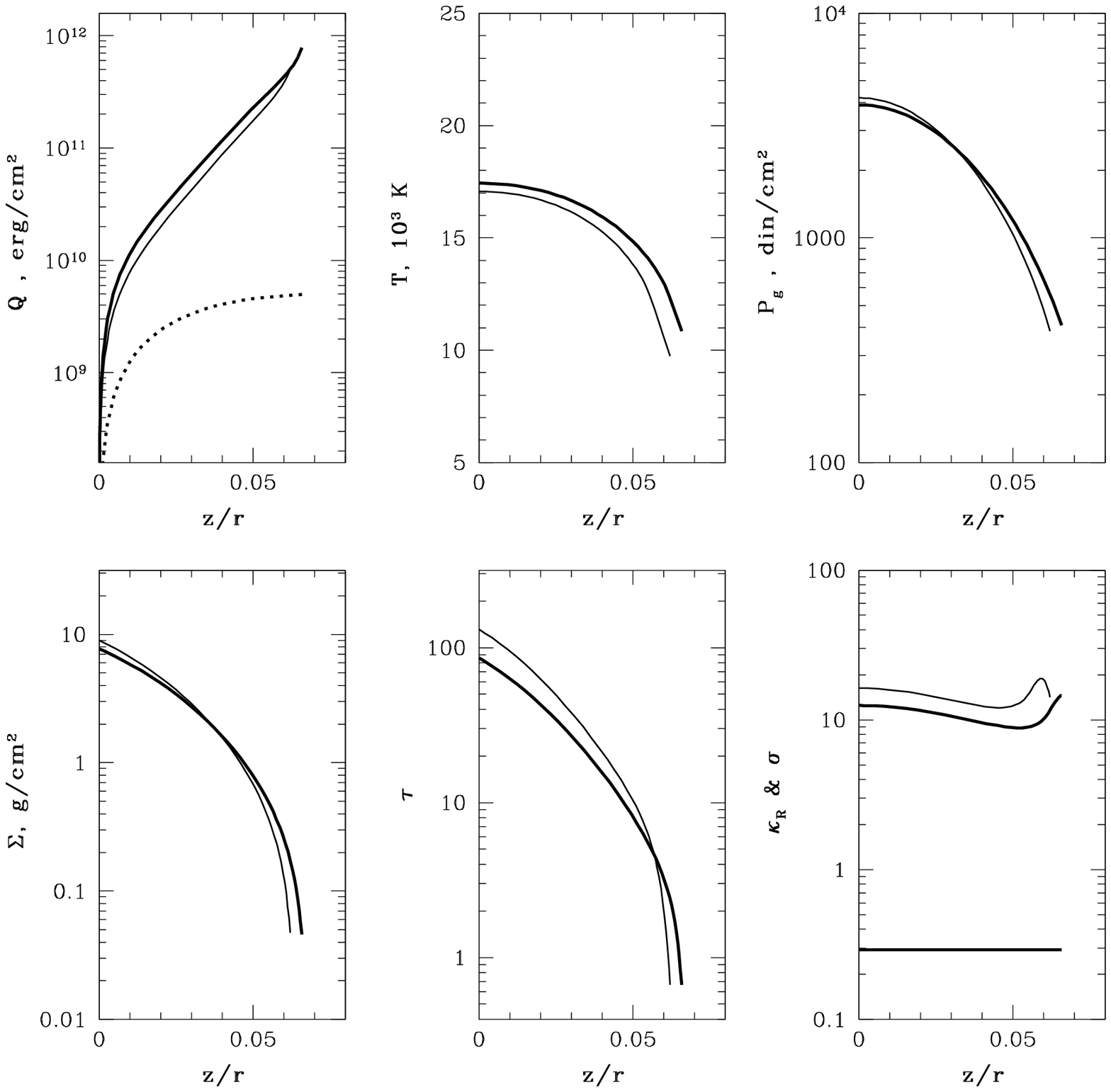}
    \caption{Vertical structure of an irradiated disk: $R=10^{11}$~см,
      $L_X=0.1\,L_{\rm Edd}$. The designations of the lines for various
      models of an irradiated disk are the same as those in Fig. 3. The
      models of an unirradiated disk and an irradiated disk without an
      atmosphere are not shown. The dotted line on the plot of $Q(z)$
      indicates the profile of the energy flux due to viscous disk heating
      in the model of a disk with an atmosphere.}
    \label{fig:risDvstrIrr} 
 \end{figure*}

Figures 3--6 present the vertical structure of an irradiated accretion disk
(the dependences $Q(z)$, $T(z)$, $P(z)$, $\Sigma(z)$, $\tau^r(z)$ and
$\kappa^r(z)$) for two radii, $R=10^{10}$\,cm and $10^11$\,cm, and two
luminosities of the central Xray source, $L_X=0.1$ and ,$1.0\,L_{\rm Edd}$. The
plots show the following models.
\begin{enumerate}
\item the irradiation of a disk with a completely scattering atmosphere
  ($\kappa_{atm}=0$) (thick solid line);
\item the irradiation of a disk with an atmosphere where absorption is
  approximately equal to scattering for photons with energy
  $E\lesssim10$\,keV (see Eq. (14); thin solid line);
\item the irradiation of a disk without an atmosphere (dashed line);
\item an unirradiated disk (dotted line).
\end{enumerate}
We considered only the models of a stationary accretion disk, i.e., those
satisfying the following condition: the temperature on the disk photosphere
should be higher than the hydrogen recombination limit.  For this reason, we
did not provide the model of an unirradiated disk for the radius
$R=10^{11}$\,cm and the calculation of an irradiated disk with an atmosphere
is shown only for $L_X=L_{\rm Edd}$. On the plots of $Q(z)$ at $R=10^{11}$\,cm,
instead of model (4) the dotted line indicates the vertical profile of the
energy flux due to viscous heating in the disk, $Q_{vis}(z)$, for a completely
scattering atmosphere.

Note that the heating of the accretion disk through irradiation at
$R=10^{10}$\,cm affects up to half of its height. At $R=10^{11}$\,cm,
irradiation can have an effect on the entire vertical disk structure, up to
the central plane. The accretion disk is deeply heated by the scattered
X-ray photons, while the direct photons from the central source are absorbed
and thermalized in a thin near-surface layer. The heating effect is enhanced
significantly in the presence of an extended atmospheric layer above the
accretion disk.

Mostly hard X-ray photons, which can traverse a sufficient distance in z
coordinate without absorption, are involved in heating the deep layers of
the accretion disk. Indeed, as follows from Eq. (17), the mean intensity of
X-ray photons at a sufficiently large depth $\Sigma$ from the surface of a
semi-infinite layer is determined by the field of scattered photons:
\be 
J(E;\Sigma)\propto C(E) \exp(-\Sigma/\Sigma_{dif}(E)) ~.
\label{Js_prop}
\ee 
Figure 7b shows the dependence of the quantity $C$ on X-ray photon energy
$E$ and, for comparison, the spectrum of the radiation incident on the disk
$S(E)$.  Figure 7a shows the characteristic surface density $\Sigma_{dif}$
(the intensity of scattered radiation with photon energy $E$ at this depth
decreases by a factor of $e$; see (34)) as a function of $E$. The quantity
$\Sigma_{dif}$ was calculated for the characteristic angle of incidence of
the X-ray photons $\cos\theta_0=0.1$ using an energy dependence of the
absorption coefficient like that for a cold gas (Morrison and McCammon
1983).
  
 \begin{figure}
   \centering 
   \includegraphics[width=1.0\linewidth]{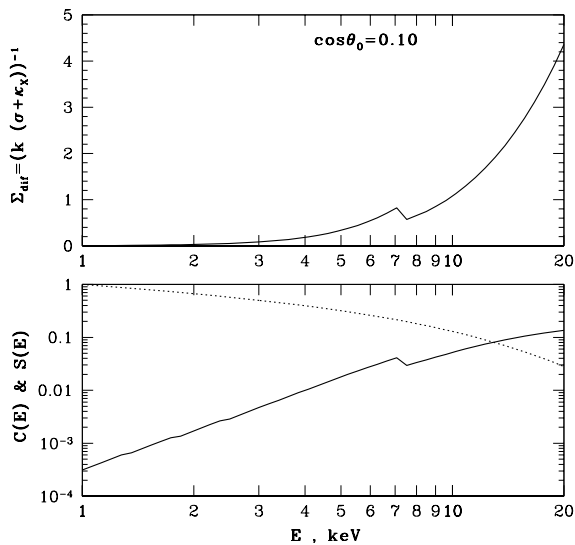}
   \caption{Energy dependence of the characteristic penetration depth of
     scattered X-ray photons $\Sigma_{dif}(E)$ (top panel) and the quantity
     $C(E)$ (bottom panel) (a semi-infinite layer, $cos\theta=0.1$). For
     comparison, the dotted line indicates the incident X-ray spectrum.}
   \label{fig:ris_Sdif_C} 
 \end{figure}

It follows from Fig. 7 that the mean intensity of scattered X-ray photons
with energy E=$10\div20$~keV is attenuated by a factor of e at a depth
$\Sigma_{dif}\approx1\div4$~g\,cm$^-2$ (which corresponds to an optical
depth in the disk for intrinsic radiation $\tau^r\approx10\div40$).
Therefore, the hard X-ray radiation from the central source (with energy
$E\gtrsim10$~keV for a cold disk) at the outer disk radius, where the total
surface density in the vertical direction reaches
$\Sigma_0\sim20$~g~cm$^-2$, is capable of heating all disk layers up to the
central plane (see Figs. 5 and 6). At the same time, the accretion disk at
this radius still remains optically thick, $\tau^r>100$, for intrinsic
radiation.

 \begin{figure*}
   \centering 
    \includegraphics[width=0.48\linewidth]{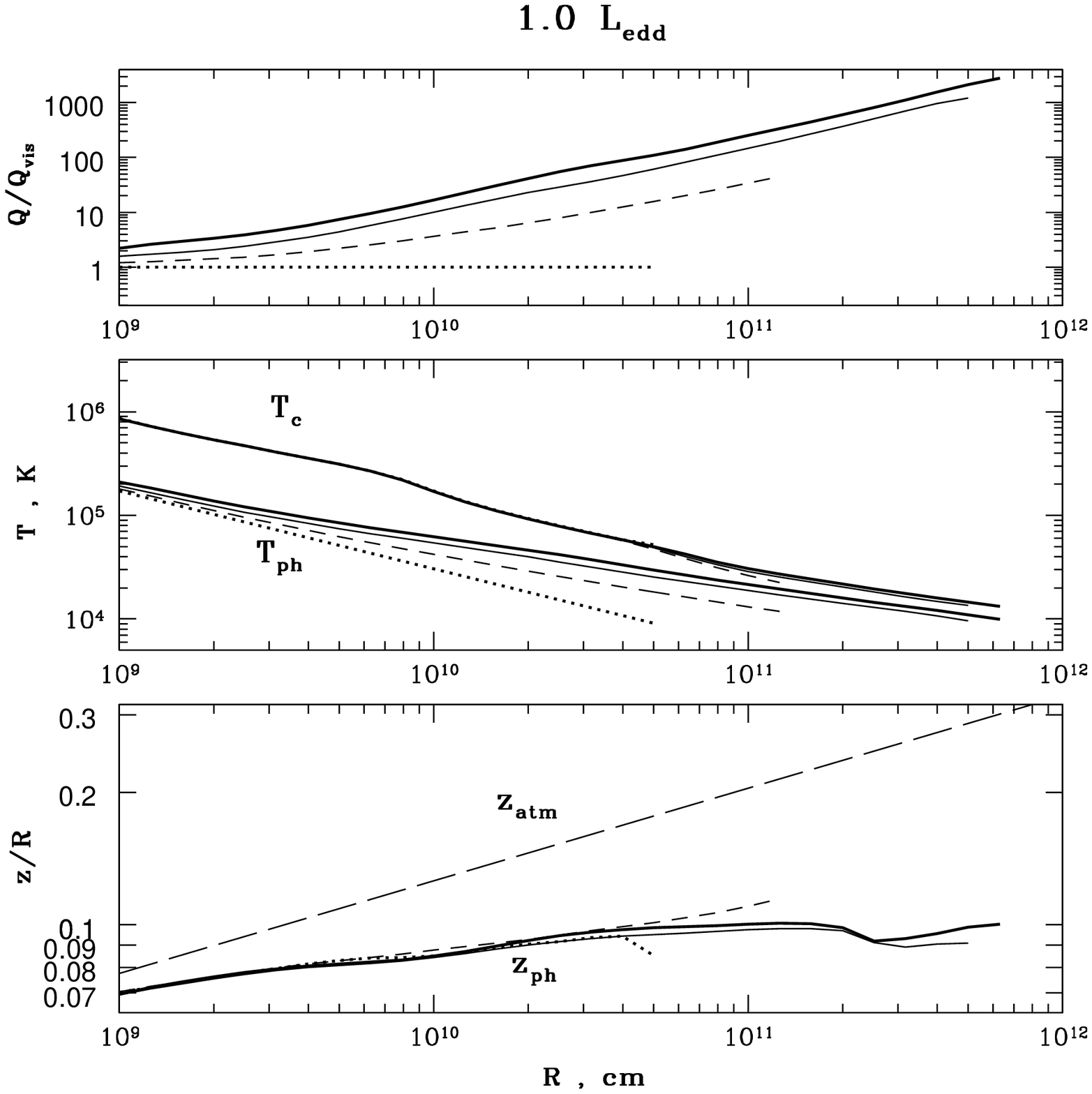}
    \includegraphics[width=0.48\linewidth]{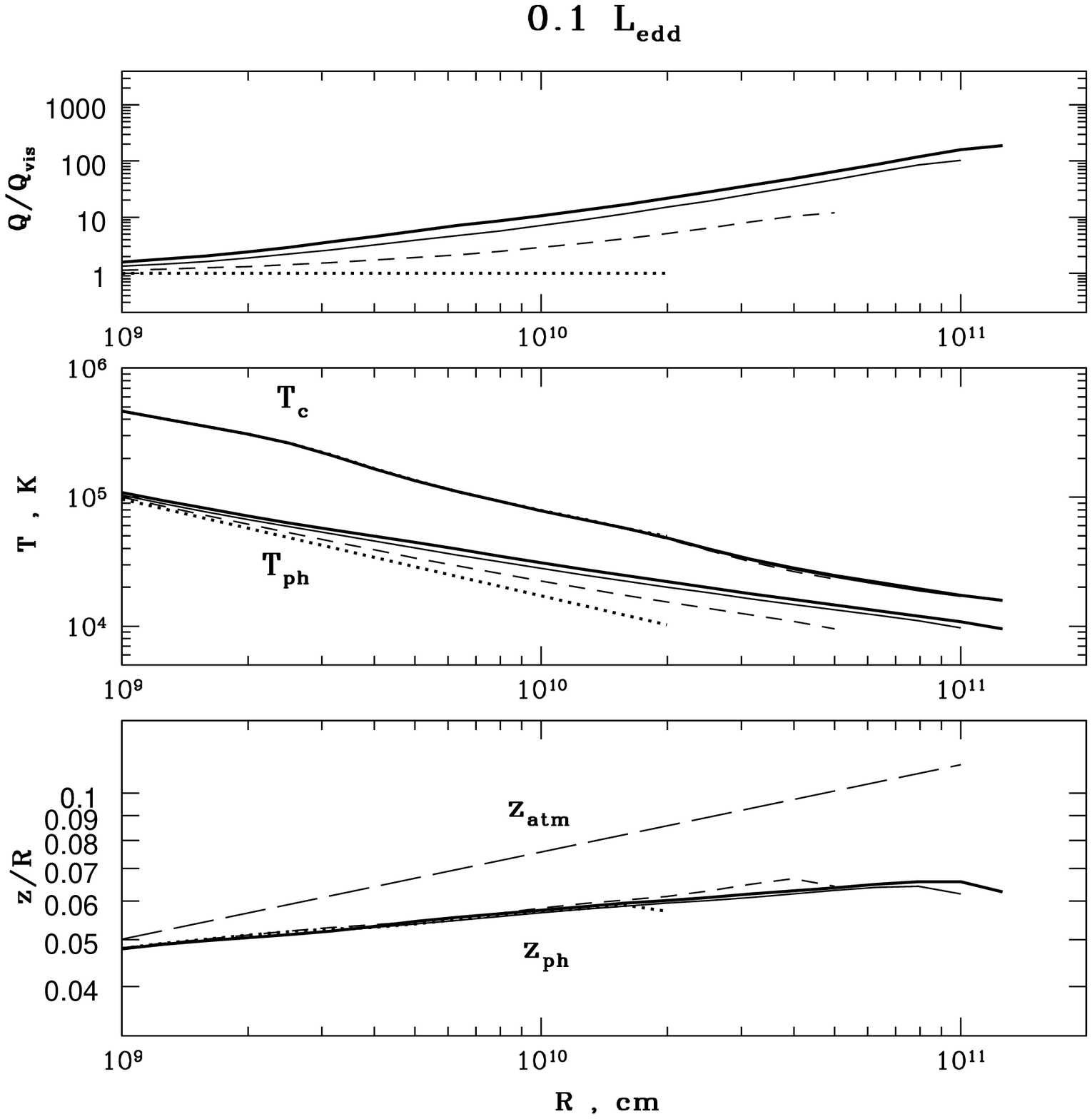} 
    \caption{Radial structure of an irradiated accretion disk for two
      luminosities of the central X-ray source: $L_X=L_{\rm Edd}$ (left
      panels) and $L_X=0.1\,L_{\rm Edd}$ (right panels). The upper row of
      plots shows the ratio of the total energy flux in the vertical
      direction to the energy flux only due to viscous heating $Q/Q_{vis}$
      at the photospheric level as a function of radius $R$; the middle
      rowsh ows the temperature in the central plane $T_c$ and on the
      photosphere $T_{ph}$ of the disk; the lower row shows the height of
      the photosphere $z_{ph}/R$ and atmosphere $z_{atm}/R$ (taken from
      Jimenez-Garate et al, 2002; the line with long dashed on the plots).
      The designations of the lines for various models of an irradiated disk
      are the same as those in Fig.~3.}
   \label{fig:ris_rstrIrr} 
 \end{figure*}

In Fig. 8, the quantities $Q/Q_{vis}$, $T_c$, $T_{ph}$ and $z_{ph}$ (the
ratio of the total flux to the viscous one at the photospheric level, the
temperature in the central plane of the disk and on the photosphere, and the
height of the photosphere, respectively) are plotted against radius $R$ for
various models of an irradiated disk. The radial profiles are shown up to
the radius at which a zone with incomplete hydrogen ionization appears in
the disk.

We will note the following:
\begin{itemize}
\item the difference between the temperatures in the central plane and on
  the photosphere of an irradiated accretion disk decreases appreciably with
  increasing radius: from $T_c/T_{ph}\approx4.5$ at $R=10^{9}$~cm to
  $T_c/T_{ph}\approx1.5$ at $R=10^{11}$~см. At $R>10^{11}$\,cm, the vertical
  structure of an optically thick disk is essentially isothermal.
\item Even strong irradiation has no significant effect on the disk
  thickness at the photospheric level $z_{ph}$ up to the radius where
  hydrogen recombination begins (and, accordingly, $z_{ph}$ decreases
  sharply). 
\item In the case of disk irradiation, the zone with incomplete hydrogen
  ionization is greatly shifted toward large radii: $R_K$ increases by a
  factor of $\sim10$ in an irradiated disk (with an atmosphere) compared to
  an unirradiated one.
\item For the luminosity of the central source $L_X=L_{\rm Edd}$, the outer
  radius of a stationary irradiated disk (with an atmosphere) can be
  $R\approx6\times10^{11}$\,cm, in qualitative agreement with the estimates
  of the outer radius in long-period persistent LMXBs.
\item In the model of an irradiated disk with an atmosphere, the problem of
  heating the outer ($R>R_K$) cold accretion disk that is screened (see
  Dubus et al. 1999) from the direct photons of the central Xray source is
  naturally resolved.
\end{itemize} 
  
\section{CONCLUSIONS}
\label{sec:conclusions}
Here, we investigated the vertical structure of the outer accretion disk in
LMXBs by taking into account the possibility of the scattering of X-ray
photons from the central source in the disk and the atmospheric layer. The
atmospheric parameters were taken from Jimenez-Garate et al. (2002). We
reached the following conclusions.
\begin{enumerate}
\item We derived simple analytic expressions for the disk heating by
  scattered X-ray photons using an approximate solution of the transfer
  equation by the Sobolev method. This approximation has a $\lesssim10\%$
  accuracy in the range of X-ray photon energies $E<20$\,keV.

\item We showed that the scattering of X-ray photons by free electrons
  affects significantly the heating of the outer accretion disk in LMXBs.
  Having been scattered, the X-ray photons with energy $E\gtrsim10$\,keV
  incident from the central source at a small angle to the disk surface can
  penetrate fairly deep into the disk photosphere and can affect
  significantly the vertical structure of the accretion disk at outer radii.

\item The scattering of X-ray photons is particularly important at large
  radii, where the total disk surface density is fairly low
  $\Sigma_0\lesssim20$~g~cm$^{-2}$ (in this case, the disk still remains
  optically thick, $\tau^r\gtrsim100$). At radii $R\gtrsim10^{11}$~cm,
  irradiation can heat all layers of an optically thick disk and its
  vertical structure becomes essentially isothermal.

\item When a disk with an atmosphere is irradiated, the radius at which a
  zone with incomplete hydrogen ionization appears in the disk increases by
  a factor of $\sim10$ compared to an unirradiated disk. For example, for
  the luminosity of the central source $L_X=L_{\rm Edd}$, the outer radius
  of a stationary irradiated disk can be $R\approx6\times10^{11}$\,cm, in
  agreement with the estimates of the outer disk radius in the long-period
  persistent LMXBs GX 13+1 and Cyg X-2.

\item In the model of an irradiated accretion disk with an atmosphere, the
  problem of the heating of its outer cold ($T<10^4$\,K) parts, which are
  screened (see Dubus et al. 1999) from the direct photons of the central
  X-ray source, is naturally resolved.
\end{enumerate}

\paragraph{ACKNOWLEDGMENTS}
This work was supported by the Program for Support of Leading Scientific
Schools of the Russian President (NSh-5069.2010.2), the Russian Foundation
for Basic Research (project nos. 09-02-00032, 10-02-00492, 10-02-91223-CTa,
09-02-97013-p-povolzh'e-a), and Programs P-19 and OFN-16 of the Russian
Academy of Sciences. N.~I.~Shakura thanks the Max Planck Institute (Germany)
for the invitation to visit this institute.
   
\section{REFERENCES}
~\\
1. C. W. Allen, Astrophysical Quantities (Athlone
Press, Univ. London, 1973).\\
2. N. R. Badnell, M. A. Bautista, K. Butler, et al., Mon.
Not. R. Astron. Soc. 360, 458 (2005).\\
3. P. B. Bosma andW. A. Rooij, Astron. Astrophys. 126,
283 (1983).\\
4. S. Chandrasekhar, Radiative Transfer (Clarendon,
Oxford, 1950).\\
5. G. Dubus, J.-P. Lasota, H.-M. Hameury, and
P. Charles, Mon. Not. R. Astron. Soc. 303, 139
(1999).\\
6. A. A. Esin, E. Kuulkers, J. E. McClintock, and
R. Narayan, Astrophys. J. 532, 1069 (2000).\\
7. M. Gilfanov and V. Arefyev, astro-ph/0501215
(2005).\\
8. M. A. Jimenez-Garate, J. C. Raymond, and
D. A. Liedahl, Astrophys. J. 581, 1297 (2002).\\
9. J. de Jong, J. van Paradijs, and T. Augusteijn, Astron.
Astrophys. 314, 484 (1996).\\
10. Y.-K. Ko and T. R. Kallman, Astrophys. J. 431, 273
(1994).\\
11. Q. Z. Liu, J. van Paradijs, and E. P. J. van den Heuvel,
Astron. Astrophys. 368, 1021 (2001).\\
12. V.M. Lyutyi and R. A. Sunyaev, Sov. Astron. 20, 290
(1976).\\
13. F. Meyer and E. Meyer-Hofmeister, Astron. Astrophys.
106, 34 (1982).\\
14. R. Morrison and D. McCammon, Astrophys. J. 270,
119 (1983).\\
15. D. I. Nagirner, Lectures on the Theory of Radiative
Transfer (SPb.Univ., St.-Petersbourg, 2001) [in
Russian].\\
16. L. A. Pozdnyakov, I. M. Sobol, and R. A. Syunyaev,
Astrophys. Space Phys. Rev. 2, 189 (1983).\\
17. W. H. Press, S. A. Teukolsky, W. T. Vetterling, and
B. P. Flannery, Numerical Recipes in FORTRAN.\\
The Art of Scientific Computing (Cambridge Univ.,
Cambridge, 1992).\\
18. D. Psaltis, Astrophys. J. 574, 306 (2002).\\
19. J. C. Raymond, Astrophys. J. 412, 267 (1993).\\
20. N. I. Shakura and R. A. Sunyaev, Astron. Astrophys.
24, 337 (1973).\\
21. V. V. Sobolev, Radiative Energy Transfer in Stellar
and Planetary Atmospheres (GITTL, Moscow,
1956) [in Russian].\\
22. V. V. Sobolev, Sov. Astron. 12, 420 (1968).\\
23. V. Suleimanov, F. Meyer, and E. Meyer-Hofmeister,
Astron. Astrophys. 350, 63 (1999).\\
24. V. Suleimanov, F. Meyer, and E. Meyer-Hofmeister,
Astron. Astrophys. 401, 1009 (2003).\\
25. V. F. Suleimanov, G. V. Lipunova, and N. I. Shakura,
Astron. Rep. 51, 549 (2007).\\
26. V. F. Suleimanov, G. V. Lipunova, and N. I. Shakura,
Astron. Astrophys. 491, 267 (2008).\\
27. Y. Tuchman, S. Mineshige, and J. C.Wheeler, Astrophys.
J. 359, 164 (1990).\\
28. S. D. Vrtilek, J. C. Raymond, M. R. Garcia, et al.,
Astron. Astrophys. 235, 162 (1990).

\clearpage
\eject

\appendix
\setcounter{equation}{0}
\renewcommand{\theequation}{П\arabic{equation}}
\section{ANALYTIC SOLUTION OF THE PROBLEM OF X-RAY PHOTON TRANSFER IN A
  PLANE-PARALLEL LAYER IN THE SOBOLEV APPROXIMATION}
\label{sec:sobolev_method}
\subsection{ A Layer with a Constant Absorption Coefficient (a Disk without an Atmosphere) }
\label{sec:sobolev_method_disk}
Consider a plane-parallel layer of material with a constant (in depth)
absorption coefficient $\kappa$ on which a parallel X-ray flux is incident
at an angle $\theta_0$ to the inward normal to the surface. The same flux is
symmetrically incident on the lower surface at an angle
($180^\circ-\theta_0$). The azimuth of the incident radiation is zero,
$\phi_0=0$.

Consider the irradiation of a cold layer with temperature $T$ by X-ray
photons with frequency $\nu$, so that the inequality $k\,T\ll h\nu$ is
valid; there are no intrinsic energy sources in the layer. We assume the
scattering in the medium to be coherent (Thomson scattering).  In this case,
the intensity of scattered radiation at frequency $\nu$,
$I(\tau,\zeta,\phi)$, satisfies the integro-differential equation (see,
e.g., Chandrasekhar 1950)
\be
\zeta \frac{\partial I(\tau,\zeta,\phi)}{\partial \tau} = -I(\tau,\zeta,\phi) + S(\tau,\zeta,\phi) ,
\label{eq:XeqRT}
\ee
The source function $S(\tau,\zeta,\phi)$ is
\be
\label{eq:Xsourcef}
\begin{array}{lll}
S(\tau,\zeta,\phi) & = & S_{inc}(\tau,\zeta,\phi) + \\
& + & \frac{\lambda}{4\pi}\int_{-1}^{+1}\int_{0}^{2\pi}x(\zeta,\phi,\zeta',\phi')I(\tau,\zeta',\phi')d\zeta'd\phi' ,
\end{array}
\ee 
where $\tau$ is the optical depth in the vertical direction measured from
the surface deep into the layer, $\zeta=\cos{\theta}$,
$\lambda=\frac{\sigma}{\kappa+\sigma}$ is the single-scattering albedo,
$\kappa$ is the absorption coefficient at frequency $\nu$, $\sigma$ is the
scattering coefficient, and $x(\zeta,\phi,\zeta',\phi')$ is the phase
function.

In the case of randomly oriented scatterings, the phase function is a
function of the scattering angle alone, $x(\cos{\gamma})$,
\be
\cos{\gamma} = \zeta\zeta' + \sqrt{1-\zeta^2}\sqrt{1-\zeta'^2}\cos{(\phi-\phi')} .
\label{eq:XcosG}
\ee 
We consider the scattering by free electrons with the phase function
\be
x(\cos{\gamma}) = \frac{3}{4}\bigg( 1 + \cos^2\gamma\bigg) = 1 + x_2 P_2(\cos{\gamma}),
\label{Xrlscat}
\ee
where $P_2(\cos{\gamma})=\frac{3\cos^2{\gamma}-1}{2}$ is the second Legendre
polynomial. The coefficient $x_2=1/2$ for the Rayleigh phase function and
$x_2=0$ for isotropic scattering.

The term $S_{inc}$ in the source function (A2) allows for the
single-scattering of the direct radiation incident on the layer:
\be
S_{inc}(\tau,\zeta,\phi) = \frac{\lambda}{4\pi}
\int_{-1}^{+1}\int_{0}^{2\pi}x(\cos\gamma)I_{inc}(\tau,\zeta',\phi')d\zeta'd\phi' ,
\ee
where $I_{inc}(\tau,\zeta,\phi)$ is the intensity of the direct radiation
in the layer::
\begin{eqnarray}
  \nonumber
  I_{inc}(\tau,\zeta,\phi) & = & F_X\,e^{-\tau/\zeta_0}\delta(\zeta-\zeta_0)\delta(\phi) \\
  & &  +  F_X\,e^{-(\tau_0 - \tau)/\zeta_0}\delta(\zeta+\zeta_0)\delta(\phi) ,
\end{eqnarray}
$\zeta_0=\cos{\theta_0}$, $\delta(x)$ is the Dirac delta function, and $F_X$
is the monochromatic illumination of the area perpendicular to the direction
of incidence of the X rays.  Thus,
\be
S_{inc}(\tau,\zeta,\phi) =  \frac{\lambda F_X}{4\pi}x(\cos{\gamma_0})
\bigg[ e^{-\tau/\zeta_0} + e^{-(\tau_0 - \tau)/\zeta_0}\bigg] ,
\label{eq:XsourcefI}
\ee
where
\be
\cos{\gamma_0} = \zeta\zeta_0 + \sqrt{1-\zeta^2}\sqrt{1-\zeta_0^2}\cos{\phi} ~.
\label{eq:XcosG0}
\ee

For an approximate solution of the transfer equation for scattered photons
(A1), (A2), (A7), we use the method developed by V.~V.~Sobolev for the
problem of diffuse reflection and transmission (Sobolev 1956). This
approximate method allows fairly simple formulas to be derived for the mean
intensity of X-ray photons in a layer $J$ and the albedo $A$ (see below and
the textbook by Nagirner 2001).

The intensity $I(\tau,\zeta,\phi)$ and the source function
$S(\tau,\zeta,\phi)$ can be written as a decomposition into azimuthal
harmonics. For the Rayleigh phase function, we have
\be
I(\tau,\zeta,\phi) = I_0(\tau,\zeta) + 2\,I_1(\tau,\zeta)\,\cos{\phi} 
+ 2\,I_2(\tau,\zeta)\,\cos{2\phi} ,
\label{eq:I_expan_phi}
\ee

\be
S(\tau,\zeta,\phi) = S_0(\tau,\zeta) + 2\,S_1(\tau,\zeta)\,\cos{\phi} + 2\,S_2(\tau,\zeta)\,\cos{2\phi} .
\ee
The mean intensity and flux of scattered X-ray photons in the vertical
direction are defined by the zeroth azimuthal harmonic of the intensity
$I_0(\tau,\zeta)$:
\begin{eqnarray}
  \nonumber
  J(\tau) & = & \frac{1}{4\pi}
  \int_{-1}^{+1}\int_{0}^{2\pi}I(\tau,\zeta,\phi)d\zeta d\phi \\
  & = & \frac{1}{2}\int_{-1}^{+1}I_0(\tau,\zeta)d\zeta ,
\end{eqnarray}
\begin{eqnarray}
  \nonumber
  H(\tau) & = &
  \int_{-1}^{+1}\int_{0}^{2\pi}\zeta I(\tau,\zeta,\phi)d\zeta d\phi \\
  & = & 2\pi\int_{-1}^{+1}\zeta I_0(\tau,\zeta)d\zeta .
\end{eqnarray}

The transfer equation for $I_0(\tau,\zeta)$ is
\be
\zeta\frac{\partial I_0(\tau,\zeta)}{\partial\tau} = -I_0(\tau,\zeta) + S_0(\tau,\zeta) .
\label{eq:XeqRT0}
\ee
We use the Eddington approximation
\begin{eqnarray}
  \nonumber
  \frac{1}{2}\int_{-1}^{+1}\zeta^2I_0(\tau,\zeta)d\zeta & = &\mu_{\rm Edd} J ,\\
  \mu_{\rm Edd} &=& \frac{1}{3} ,
\label{muEdd}
\end{eqnarray}
where $\mu_{\rm Edd}$ is the constant Eddington factor. In the general (more
accurate) case, $\mu_{\rm Edd}$ changes with $z$ coordinate.

Substituting (A3), (A4), (A7), and (A9) into (A2), using the Eddington
approximation (A14), and gathering the azimuth-independent terms, we have
for the source function
\begin{eqnarray}
  \nonumber 
  S_0(\tau,\zeta) & = & \lambda J + 
\frac{\lambda F_X}{4\pi}
\bigg[ 1 + \cfrac{x_2}{4}\big(1-3\zeta_0^2\big)\big(1-3\zeta^2\big)\bigg]\times \\
& & \times \bigg[ e^{-\tau/\zeta_0} + e^{-(\tau_0 - \tau)/\zeta_0}\bigg] ,
\label{eq:Xsourcef0} 
\end{eqnarray}
$x_2=1/2$ in the case of scattering with the Rayleigh phase function and
$x_2=0$ in the case of isotropic scattering.

Let us integrate the transfer equation for the zeroth
harmonic (A13) over $\zeta$ with the weight factors 1
and $\zeta$. Applying again the Eddington approximation,
we will obtain the following system of differential
equations for the functions $J$ and $h=\frac{H}{4\pi}$:
\be
\label{eq:dh}
\frac{d h}{d \tau} = - (1 - \lambda) J + \frac{\lambda F_X}{4\pi} 
\bigg[ e^{-\tau/\zeta_0} + e^{-(\tau_0 - \tau)/\zeta_0} \bigg] ~,
\ee
\be
\label{eq:h}
h = -\frac{1}{3}\frac{d J}{d \tau}  ~.
\ee

Substituting $h$ from (A17) into (A16), we will obtain an inhomogeneous
second-order differential equation for the monochromatic mean intensity of
scattered photons:
\be
\frac{d^2 J}{d \tau^2} - k^2 J = - \frac{3 \lambda F_X}{4 \pi} 
\bigg[ e^{-\tau/\zeta_0} + e^{-(\tau_0 - \tau)/\zeta_0} \bigg] ,
\label{eq_J}
\ee
where $k = \sqrt{3(1-\lambda)}$. Having determined $J(\tau)$, we can find
the flux of scattered photons $H(\tau)$ from Eq. (A17):
\be
\label{h4pi}
H = -\frac{4\pi}{3}\frac{d J}{d \tau} .  
\ee 

Interestingly, Eqs. (A18) and (A19) derived by the Sobolev method have the
same form for the isotropic or Rayleigh phase function (the parameter $x_2$
does not enter into the formulas).

A general solution of the inhomogeneous differential equation (A18) should
be sought in the form of the sum
\be
J(\tau) = J_p(\tau) + J_g(\tau) ,
\label{J_sform}
\ee
where $J_p(\tau)$ is a particular solution of the inhomogeneous equation
(A18) and $J_g(\tau)$ is the general solution of the homogeneous equation
\be 
\frac{d^2 J}{d \tau^2} - k^2 J = 0 .
\label{eq_J_ode}
\ee

The particular solution of the inhomogeneous differential
equation (A18) is
\be
J_p = - \frac{D\,F_X}{4 \pi}
\bigg( e^{-\tau/\zeta_0} + e^{-(\tau_0 - \tau)/\zeta_0} \bigg) .
\label{J_p}
\ee
The coefficient $D$ can be found by directly substituting
solution (A22) into Eq. (A18):
\be
\label{eqD}
 D = \frac{3 \lambda \zeta_0^2}{1 - k^2\zeta_0^2}.\\
\ee 

The general solution of the homogeneous equation (A21) is
\be
J_g(\tau) = \frac{ F_X}{4 \pi} \times \left\{ 
\begin{array}{ll}
C_1 e^{-k\tau} + C_2 e^{k\tau} & , \lambda<1, \kappa>0 \\
C_1 \tau       + C_2           & , \lambda=1, \kappa=0
\end{array} 
\right.
\label{J_g}
\ee
We are interested in the case of $\kappa>0$, $\lambda<1$.

Since the problem is symmetric relative to the $\tau=\tau_0/2$ plane, we
will obtain
\be
J_g(\tau) = \frac{ C\, F_X}{4 \pi} \bigg[e^{-k\tau} + e^{-k(\tau_0-\tau)}\bigg] .
\label{J_g1}
\ee

To find the coefficient $C$, we use the condition at the outer boundary of
the layer $\tau=0$:
\be
\bigg( J - \frac{2}{3} \frac{d J}{d \tau} \bigg) \Bigg|_{\tau=0} = 0 .
\label{bc1}
\ee
This is equivalent to $H\big|_{\tau=0}=-2\pi J\big|_{\tau=0}$ --- there is
no externally incident scattered radiation at the boundary of the layer.
Substituting (A20) into condition (A26) yields an expression for the
coefficient $C$:
\be
  C = D \times
  \cfrac{1+e^{-\tau_0/\zeta_0} + \cfrac{2}{3\zeta_0}\bigg(1+e^{-\tau_0/\zeta_0}\bigg)}
  {1+e^{-k\tau_0} + \cfrac{2k}{3}\bigg(1+e^{-k\tau_0}\bigg)} .
\label{eqC}
\ee

Thus, the mean intensity and flux of scattered Xray photons in the layer at
an optical depth $\tau$ are
\begin{eqnarray}
  \nonumber
  J & = & 
   \frac{F_X}{4\pi}C\bigg[e^{-k\tau} + e^{-k(\tau_0-\tau)}\bigg] -\\
  & & \frac{F_X}{4\pi}D\bigg[ e^{-\tau/\zeta_0} + e^{-(\tau_0 - \tau)/\zeta_0} \bigg] ,
  \label{Js}
\end{eqnarray}
\begin{eqnarray}
  \nonumber
  H & = & 
   F_X\frac{k\,C}{3}\bigg[e^{-k\tau} - e^{-k(\tau_0-\tau)}\bigg] - \\
  & & F_X\frac{D}{3\zeta_0}
  \bigg[ e^{-\tau/\zeta_0} - e^{-(\tau_0 - \tau)/\zeta_0} \bigg]  ,
  \label{Hs}
\end{eqnarray}
where the coefficients $D$ and $C$ are determined from Eqs. (A23) and (A27),
respectively.

We emphasize that $D$ is not an independent coefficient; it is related to
the coefficient $C$ (see (A27)). We are dealing with a second-order
differential equation that has two independent solutions. For the
semiinfinite problem, the second solution is ruled out, because it grows
exponentially. For a layer with a finite optical depth ($\tau_0$), we
naturally have a solution with two constants $C_1$ and $C_2$ that, in our
case, turn into one constant $C$, because the upper and lower parts of the
layer are irradiated symmetrically.

The above expressions for the mean intensity $J(\tau)$ and flux $H(\tau)$ of
scattered photons in the layer are so structured that we have an
indeterminate form of the type ($\infty-\infty$) at $k\zeta_0=1$,
$\lambda=\lambda_\star$,
\be
\lambda_\star = 1-\frac{1}{3\zeta_0^2} ~,  
\ee
because the denominator of the coefficients $D$ and $C$ becomes zero. Let us
evaluate this indeterminate form. At $\lambda=\lambda_\star$, the functions
$J(\tau)$ and $H(\tau)$ will then have finite values:
\be
J(\lambda_\star;\tau) =
   \frac{F_X}{4\pi} (C-D)_{k=\frac{1}{\zeta_0}}
\bigg( e^{-\tau/\zeta_0} + e^{-(\tau_0 - \tau)/\zeta_0} \bigg)
\label{Js_kzeta1}
\ee
\be
H(\lambda_\star;\tau) =
   \frac{F_X}{3\zeta} (C-D)_{k=\frac{1}{\zeta_0}}
\bigg( e^{-\tau/\zeta_0} - e^{-(\tau_0 - \tau)/\zeta_0} \bigg) ,
\label{Hs_kzeta1}
\ee
\be
(C-D)_{k=\frac{1}{\zeta_0}} =
\cfrac{\lambda_\star\zeta_0}{1+\frac{2}{3\zeta_0}} \times
 \cfrac{1+\bigg[1 - \frac{3\tau_0}{2}\bigg(1 + \frac{2}{3\zeta_0}\bigg)\bigg]e^{-\tau_0/\zeta_0}}{1+e^{-\tau_0/\zeta_0}}
\label{C-D_kzeta1}
\ee
If the X-ray photons are incident at a fairly small angle
$(90^\circ-\theta_0)$ to the layer surface, more specifically,
$(90^\circ-\theta_0)<arcos(1/\sqrt{3})\approx35.3^\circ$, then the condition
$k\zeta_0<1$ is always met and the denominator of the coefficient $D$ does
not become zero for any $\lambda$.

It is important to note that there must also be an angle of incidence in the
exact solution of the transfer problem with external irradiation where a
similar indeterminate form emerges. It stems from the fact that at large
angles of incidence the field of scattered photons decreases with depth more
rapidly than the field of incident photons, while at small angles the field
of incident photons decreases more rapidly.

Thus, the Eddington approximation ($\mu_{\rm Edd}=1/3$, see (A14)) gives a
qualitatively newresu lt (the appearance of an indeterminate form at some
critical angle of incidence) in problems with external irradiation.  As the
critical angle is approached, the constants $C$ and $D$ tend to infinity,
but their difference $C-D$ remains finite. Therein lies the deep meaning of
the approximate (Sobolev) solution. This result must also be retained in the
exact solution of the problem.

The mean intensity and flux in the vertical direction of the direct
radiation incident on the layer are
\begin{eqnarray}
J_{inc} & = & F_X\Big( e^{-\tau/\zeta_0}+e^{-(\tau_0-\tau)/\zeta_0} \Big) ,\\
\label{Jinc}
H_{inc} & = & \zeta_0 F_X\Big( e^{-\tau/\zeta_0}-e^{-(\tau_0-\tau)/\zeta_0} \Big) .
\label{Hinc}
\end{eqnarray}
The total mean intensity and flux of X-ray photons in the layer at an
optical depth $\tau$ (direct and scattered radiation) are
\begin{eqnarray}
  \nonumber
  J_{tot} & = & 
  \frac{F_X}{4\pi}C\bigg[e^{-k\tau} + e^{-k(\tau_0-\tau)}\bigg] + \\
 & & \frac{F_X}{4\pi}(1-D)
  \bigg[ e^{-\frac{\tau}{\zeta_0}} + e^{-\frac{\tau_0 - \tau}{\zeta_0}} \bigg] ,
  \label{eq:Jtot}  
\end{eqnarray}
\begin{eqnarray}
 \nonumber
  H_{tot} & = & 
  F_X\frac{k\,C}{3}\bigg[e^{-k\tau} - e^{-k(\tau_0-\tau)}\bigg] +\\
 & & F_X\bigg(\zeta_0-\frac{D}{3\zeta_0}\bigg)
  \bigg[ e^{-\frac{\tau}{\zeta_0}} - e^{-\frac{\tau_0 - \tau}{\zeta_0}} \bigg] .
  \label{eq:Htot}
\end{eqnarray}
If the X-ray photons are incident at a small angle to the surface, then the
direct radiation does not pass deep into the layer, because the exponential
$e^{-\tau/\zeta_0}$, $1/\zeta_0\gg1$, decreases rapidly and the radiation
field inside the layer is determined by diffuse radiation.

The albedo of the layer for X-ray photons is
\be
A = - \frac{H}{H_{inc}}\bigg|_{\tau=0} .
\label{genAlbedo}
\ee
Substituting $H$ and $H_{inc}$ from (A29) and (A35) yields
\be
A = \frac{D}{3\zeta_0^2} - \cfrac{k\,C}{3\zeta_0} 
\bigg( \frac{1 - e^{-k\tau_0}}{1 - e^{-\tau_0/\zeta_0}} \bigg) .
\label{Albedo}
\ee

To estimate the accuracy of the derived formulas for themean intensity and
flux of X-ray photons (A36) and (A37), we compared the albedos (A39) with
their exact values for various angles of incidence of the Xray photons in
the case of a semi-infinite medium ($\tau_0\rightarrow\infty$). The exact
albedos for the isotropic and Rayleigh phase functions were calculated via
the Chandrasekhar $H$-functions (for more detail, see Appendix 3).

The table gives approximate albedos obtained by the Sobolev method for
various $\zeta$ and $\lambda$, and their exact values for isotropic and
Rayleigh scattering. We see excellent agreement between the exact and
approximate albedos.

\subsection{ A Plane-Parallel Layer with Two Absorption Coefficients
  (Disk+Atmosphere) } \label{sec:sobolev_method_disk_atm}

Let us divide a plane layer into three zones at the boundary of which the
absorption coefficient changes abruptly (see Fig. 1): $\kappa=\kappa_{a}$ in
zones 1 and 3 and $\kappa=\kappa_d$ in zone 2. The middle zone 2 corresponds
to a cold disk, while zones 1 and 3 correspond to atmospheric regions at the
top and the bottom. The central plane of the disk passes in the middle of
zone 2 and is the plane of symmetry of the problem. Denote the optical depth
in the vertical direction in zones 1 and 3 by $\tau_a$ and the optical depth
between the boundary of zone 2 and the plane of symmetry of the layer by
$\tau_d$ (see Fig.  1); the total optical depth in the layer is 
$\tau_0 = 2(\tau_{a} + \tau_d)$.

Inside each zone, we seek a solution of Eq. (A18) with a constant $\lambda$:
\be \lambda = \left\{
\begin{array}{ll}
 \lambda_{a} &, 0<\tau<\tau_{a} \\
 \lambda_d &, \tau_{a}<\tau<\tau_0-\tau_{a} \\
 \lambda_{a} &, \tau_0-\tau_{a}<\tau<\tau_0 .\\
\end{array} 
\right.
\label{k3}
\ee
where
$\lambda_{a} = \frac{\sigma}{\kappa_{a} +\sigma}$, 
$\lambda_d = \frac{\sigma}{\kappa_d +\sigma}$.

It will suffice to find a solution in the region $0<\tau<\tau_0/2$. Since
the problem is symmetric relative to the $\tau=\tau_0/2$ plane, the solution
for the symmetric zone $\tau_0/2<\tau<\tau_0$ can be obtained by the change
of variables $\tau\Longleftrightarrow\tau_0 -\tau$.

For $\lambda_{a} and \lambda_d<1$ ($\kappa_a,\kappa_d>0$), the general
solution of the inhomogeneous differential equation (A18) in zone 1, $0<\tau<\tau_a$:
\begin{eqnarray}
  \nonumber
  J(\tau) &=& \frac{ F_x}{4 \pi} \bigg(
  C_{1a} e^{-k_{a}\tau} + C_{2a} e^{k_{a}\tau} \bigg) - \\
  & & - \frac{ F_x}{4 \pi} D_{a}\bigg( e^{-\frac{\tau}{\zeta_0}} + e^{-\frac{\tau_0 - \tau}{\zeta_0}}
  \bigg),
\label{Jdc_c}
\end{eqnarray}
$k_{a} = \sqrt{3(1-\lambda_{a})}$.

In zone 2, $\tau_{a}<\tau<\tau_0/2$, given the
symmetry relative to the $\tau_0/2$ plane, we have
\begin{eqnarray}
  \nonumber
  J(\tau) &=& \frac{ F_x C_d}{4 \pi}
  \bigg( e^{-k_d\tau} 
  + e^{-k_d(\tau_0-\tau)} \bigg) - \\ 
  & & - \frac{ F_x D_{d}}{4 \pi} \bigg( e^{-\frac{\tau}{\zeta_0}} + 
  e^{-\frac{(\tau_0 - \tau)}{\zeta_0}} \bigg),
  \label{Jdc_d} 
\end{eqnarray}
$k_d = \sqrt{3(1-\lambda_d)}$.

The coefficients $D_a$ and $D_d$ are defined by the expressions
\begin{eqnarray}
 D_{a} & = & \frac{3 \lambda_{a} \zeta_0^2}{1-k_{a}^2\zeta_0^2},\\
\label{eqD1}
 D_{d} & = & \frac{3 \lambda_d \zeta_0^2}{1-k_d^2\zeta_0^2}. 
\label{eqD2} 
\end{eqnarray} 

To find the coefficients $C_{1a}$, $C_{2a}$, and $C_d$, it is necessary to
use the following three boundary conditions:
\begin{itemize}
\item the condition at the upper boundary of the layer (A26);
\item at the boundary of zones 1 and 2, the quantities $\kappa$, $\lambda$
  and $k$ undergo a discontinuity, while the quantities $J$ and $H$ must
  change continuously. Hence two joining conditions for the mean intensity
  and its derivative follow:
\begin{eqnarray}
\label{bc2}
 J\bigg|_{\tau=\tau_{a}-0} & = & J\bigg|_{\tau=\tau_{a}+0} ,\\
\label{bc3}
 \frac{d J}{d \tau}\bigg|_{\tau=\tau_{a}-0} & = & 
 \frac{d J}{d \tau}\bigg|_{\tau=\tau_{a}+0} .
\end{eqnarray}
\end{itemize}

Substituting (A41) and (A42) into conditions (A26), (A45), and (A46), we
obtain a system of three linear algebraic equations with three unknowns
solving which we find $C_{1a}$, $C_{2a}$, and $C_d$:
\be
C_{2a} =
\cfrac{ 
  \cfrac{D_{a} f}{1+\frac{2}{3}k_{a}} \Big[\frac{g_-}{g_+}-k_{a}\Big] e^{-k_{a}\tau_a}
  - (D_{a}-D_d)\Big[\frac{g_- b_+}{g_+} -\frac{b_-}{\zeta_0}\Big]
}{
  \cfrac{1-\frac{2}{3}k_{a}}{1+\frac{2}{3}k_{a}} \Big[\frac{g_-}{g_+}-k_{a}\Big]
  e^{-k_{a}\tau_a} - \Big[\frac{g_-}{g_+}+k_{a}\Big] e^{k_{a}\tau_a} 
} ~,
\label{eqC2a}
\ee
\be
C_{1a} = \cfrac{D_{a} f}{1+\frac{2}{3}k_{a}}
   - C_{2a} \cfrac{1-\frac{2}{3}k_{a}}{1+\frac{2}{3}k_{a}} ~,
\label{eqC1a}
\ee
\be
C_d = \cfrac{C_{1a}e^{-k_a\tau_a} + C_{2a}e^{k_a\tau_a} - (D_a-D_d)b_+}{g_+} ~.
\label{eqCd}
\ee
The auxiliary quantities $f$, $b_+$, $b_-$, $g_+$ and $g_-$ are defined by
the formulas:
\begin{eqnarray}
\nonumber
f &=& 1 + \frac{2}{3\zeta_0} + \bigg(1-\frac{2}{3\zeta_0}\bigg) 
e^{-\frac{\tau_0}{\zeta_0}} ~,\\
\nonumber
b_+ &=& e^{-\frac{\tau_a}{\zeta_0}} + e^{-\frac{\tau_0-\tau_a}{\zeta_0}} ~,\\
\nonumber
b_- &=& e^{-\frac{\tau_a}{\zeta_0}} - e^{-\frac{\tau_0-\tau_a}{\zeta_0}} ~,\\
\nonumber
g_+ &=& e^{-k_d\tau_a} + e^{-k_d(\tau_0-\tau_a)} ~,\\
\nonumber
g_- &=& k_d \times \Big( e^{-k_d\tau_a} - e^{-k_d(\tau_0-\tau_a)}\Big) ~.
\end{eqnarray}

As above, the flux of scattered photons $H(\tau)$ and the albedo of the
layer $A$ can be found using Eqs. (A19) and (A38).

\subsection{The Albedo of a Semi-infinite Layer: Exact
  Values}\label{sec:exact_inf_albedo}
The albedo of a semi-infinite atmosphere in the case of coherent scattering
can be calculated using the Chandrasekhar $H$-functions (Chandrasekhar
1950). The $H$-function can be found as a solution of the nonlinear integral
equation
\be
H(\mu) = 1 + \mu H(\mu)\int_0^1\frac{\psi(\mu)H(\eta)d\mu}{\mu + \eta},
\label{Hchandra}
\ee 
where $\psi(\mu)$ is the characteristic function (Chandrasekhar 1950) for
the chosen phase function $x(\mu)$; $0\le\mu\le1$. Consider two cases: (1)
isotropic scattering and (2) Rayleigh scattering.

In these cases, the characteristic function is (Sobolev 1968)
\be
\psi(\mu) = \frac{\lambda}{2} 
\bigg[ 1 + \frac{x_2}{2}(3(1-\lambda)\mu^2 - 1) P_2(\mu) \bigg] ,
\label{fpsi_rl}
\ee
$P_2(\mu)$ is the second Legendre polynomial, $x_2=0$ for isotropic
scattering, $x_2=1/2$ for Rayleigh scattering. The characteristic function
(A51) satisfies the condition $\int_0^1\psi(\eta)d\eta\le\frac{1}{2}$.

We found the function $H(\mu)$ using the procedure of successive iterations
proposed by Bosma and Rooij (1983) (see method 3). Having determined
$H(\mu)$, we found the albedo of a semi-infinite atmosphere $A_p(\zeta)$
from the formulas (Sobolev 1968)
\begin{eqnarray}
A_p(\zeta) &=& 1 - \frac{1-\lambda}{\Delta} H(\zeta) (N_2 - N_1\zeta) ,\\
\label{Albedo_exact}
\nonumber
\Delta &=& M_1 N_2 - M_2 N_1 ~, \\
\nonumber
M_1 &=& 1 - \frac{\lambda}{2}\int_0^1 H(\eta)\bigg[ 1 -
\frac{x_2}{2}P_2(\eta)\bigg] d\eta ~,\\
\nonumber
M_2 &=& - \frac{\lambda}{2}\int_0^1 H(\eta)\bigg[ 1 -
\frac{x_2}{2}P_2(\eta)\bigg]\eta d\eta ~,\\
\nonumber
N_1 &=& \frac{\lambda}{4}x_2 3(1-\lambda) 
\int_0^1 H(\eta)P_2(\eta)\eta d\eta ~,\\
\nonumber
N_2 &=& M_1 .
\nonumber
\end{eqnarray}
The albedos of a semi-infinite atmosphere for various $\zeta_0$ and
$\lambda$ for the isotropic and Rayleigh phase functions are given in the
table.
 
\clearpage
\eject

\label{lastpage}

 
 \begin{sidewaystable}
 \small
  \centering 
  \begin{tabular}{@{}rcccccccc@{}}
    \multicolumn{9}{c}{{\bf Table.} Albedos for a semi-infinite layer
    }\\
    \multicolumn{9}{c}{
      in the Sobolev approximation ($A_S$) and their exact values for the isotropic ($A_{iso}$) and
      Rayleigh ($A_{iso}$) phase functions.
    }\\ 
    \hline
    & $\zeta=0.01$ & $\zeta=0.05$ & $\zeta=0.10$ & $\zeta=0.20$ & $\zeta=0.40$ & $\zeta=0.60$ & $\zeta=0.80$ & $\zeta=1.00$ \\
    \hline
    $\lambda$ 
    &$A_{S}$ $A_{iso}$ $A_{rl}$
    &$A_{S}$ $A_{iso}$ $A_{rl}$
    &$A_{S}$ $A_{iso}$ $A_{rl}$
    &$A_{S}$ $A_{iso}$ $A_{rl}$
    &$A_{S}$ $A_{iso}$ $A_{rl}$
    &$A_{S}$ $A_{iso}$ $A_{rl}$
    &$A_{S}$ $A_{iso}$ $A_{rl}$
    &$A_{S}$ $A_{iso}$ $A_{rl}$
    \\ 
    \hline
    0.10 
    & 0.047 0.049 0.049
    & 0.044 0.044 0.043
    & 0.041 0.040 0.039
    & 0.036 0.034 0.033
    & 0.029 0.026 0.026
    & 0.024 0.022 0.022
    & 0.021 0.019 0.019
    & 0.018 0.016 0.018
    \\ 
    0.20 
    & 0.097 0.101 0.101
    & 0.091 0.091 0.090
    & 0.085 0.083 0.081
    & 0.075 0.071 0.069
    & 0.061 0.056 0.055
    & 0.051 0.047 0.047
    & 0.044 0.040 0.041
    & 0.039 0.035 0.038
    \\ 
    0.30 
    & 0.150 0.157 0.157
    & 0.142 0.143 0.142
    & 0.133 0.130 0.128
    & 0.118 0.112 0.110
    & 0.097 0.090 0.088
    & 0.082 0.075 0.075
    & 0.071 0.065 0.067
    & 0.062 0.057 0.061
    \\ 
    0.40 
    & 0.208 0.218 0.218
    & 0.198 0.199 0.198
    & 0.186 0.183 0.181
    & 0.166 0.159 0.157
    & 0.137 0.128 0.127
    & 0.117 0.109 0.109
    & 0.102 0.094 0.096
    & 0.090 0.083 0.088
    \\ 
    0.50 
    & 0.272 0.284 0.284
    & 0.259 0.262 0.261
    & 0.245 0.242 0.240
    & 0.221 0.213 0.210
    & 0.185 0.174 0.172
    & 0.159 0.148 0.148
    & 0.139 0.130 0.132
    & 0.124 0.115 0.121
    \\ 
    0.60 
    & 0.343 0.358 0.358
    & 0.329 0.333 0.332
    & 0.313 0.310 0.308
    & 0.284 0.276 0.273
    & 0.241 0.229 0.227
    & 0.209 0.197 0.197
    & 0.185 0.174 0.176
    & 0.165 0.155 0.161
    \\ 
    0.70 
    & 0.425 0.442 0.443
    & 0.409 0.415 0.415
    & 0.392 0.390 0.389
    & 0.360 0.352 0.350
    & 0.311 0.299 0.297
    & 0.273 0.261 0.260
    & 0.244 0.232 0.234
    & 0.220 0.209 0.214
    \\ 
    0.80 
    & 0.524 0.543 0.545
    & 0.508 0.516 0.517
    & 0.490 0.491 0.490
    & 0.457 0.451 0.449
    & 0.403 0.391 0.389
    & 0.360 0.348 0.347
    & 0.326 0.313 0.315
    & 0.297 0.285 0.290
    \\ 
    0.90 
    & 0.656 0.675 0.678
    & 0.642 0.652 0.654
    & 0.625 0.629 0.630
    & 0.594 0.592 0.591
    & 0.541 0.532 0.531
    & 0.496 0.486 0.485
    & 0.458 0.447 0.449
    & 0.426 0.415 0.418
    \\ 
    0.95 
    & 0.752 0.770 0.773
    & 0.741 0.751 0.754
    & 0.727 0.733 0.734
    & 0.701 0.701 0.701
    & 0.654 0.649 0.648
    & 0.613 0.605 0.605
    & 0.576 0.568 0.569
    & 0.544 0.536 0.537
    \\ 
    0.99 
    & 0.886 0.897 0.899
    & 0.880 0.887 0.889
    & 0.872 0.878 0.879
    & 0.858 0.860 0.861
    & 0.830 0.829 0.830
    & 0.804 0.802 0.802
    & 0.780 0.776 0.776
    & 0.756 0.753 0.753
    \\ 
    \hline
 \end{tabular} 
 \end{sidewaystable}

\end{document}